\documentclass[article,5p,times, twocolumn]{elsarticle}
\usepackage{fancyhdr}%
\newlength\FHoffset
\fancyheadoffset{\FHoffset}
\newlength\FHleft
\newlength\FHright
\newbox\FHline
\setbox\FHline=\hbox{\hsize=\paperwidth%
  \hspace*{\FHleft}%
  \rule{\dimexpr\headwidth-\FHleft-\FHright\relax}{\headrulewidth}\hspace*{\FHright}%
}

\renewcommand{\headrulewidth}{0.2pt}
\usepackage{titlesec}
\usepackage{color}
\usepackage{multirow}
\usepackage{natbib}
\usepackage{subfloat}
\usepackage{amsmath}
\usepackage{amssymb}
\usepackage{braket}
\usepackage{latexsym}
\usepackage{paralist}
\usepackage{graphics}
\usepackage{lscape}
\usepackage{a4wide}
\usepackage{longtable} %tab
\usepackage{relsize}   %tab
\usepackage{colortbl}  %tab
\usepackage{xcolor}    %tab
\usepackage{url}
\usepackage{enumitem}
\usepackage{amssymb}
\usepackage{amsthm}
\usepackage[export]{adjustbox}
\usepackage{lineno}
\usepackage[figuresright]{rotating}
\usepackage{graphicx}
\usepackage{caption}
\usepackage{subcaption}

\begin{document}

\begin{frontmatter}

\title{Design and characterization of a single photoelectron calibration system for the NectarCAM camera of the medium-sized telescopes of the Cherenkov Telescope Array}

%% use optional labels to link authors explicitly to addresses:
%\corref{corSN} 
\author[1]{Barbara Biasuzzi}
\author[1]{Kevin Pressard}
\author[1]{Jonathan Biteau}
\author[1]{Brice Geoffroy}
\author[1]{Carlos Domingues Goncalves}
\author[1]{Giulia Hull}
\author[1]{Miktat Imre}
\author[1]{Michael Josselin}
\author[1]{Alain Maroni}
\author[1]{Bernard Mathon}
\author[1]{Lucien Seminor}
\author[1]{Tiina Suomijarvi}
\author[1]{Thi Nguyen Trung}
\author[1]{Laurent Vatrinet}
%=======================================
\author[5]{Patrick Brun}           %LUPM
\author[2]{Sami Caroff\fnref{fn1}} %LLR
\author[2]{Stephen Fegan}          %LLR
\author[2]{Oscar Ferreira}         %LLR
\author[6]{Pierre Jean}            %IRAP
\author[3]{Sonia Karkar}           %LPNHE
\author[6]{Jean-Fran\c{c}ois Olive}%IRAP
\author[5]{St\'ephane Rivoire}     %LUPM
\author[4]{Patrick Sizun}          %CEA
\author[2]{Floris Thiant}          %LLR
\author[6]{Adellain Tsiahina}      %IRAP
\author[3]{Fran\c{c}ois Toussenel} %LPNHE
\author[5]{Georges Vasileiadis}    %LUPM

\fntext[fn1]{Currently affiliated to $c$.}
%=======================================
\address[1]{Institut de Physique Nucl\'eaire, IN2P3/CNRS, Universit\'e Paris-Sud, Universit\'e Paris-Saclay, 15 rue Georges Cl\'emenceau, 91406 Orsay, Cedex, France}
\address[2]{Laboratoire Leprince-Ringuet, \'Ecole Polytechnique (UMR 7638, CNRS/IN2P3, Universit\'e Paris-Saclay), 91128 Palaiseau, France}
\address[3]{Sorbonne Universit\'es, UPMC Universit\'e Paris 06, Universit\'e Paris Diderot, Sorbonne Paris Cit\'e, CNRS, Laboratoire de Physique Nucl\'eaire et de Hautes Energies (LPNHE), 4 Place Jussieu, 75252, Paris Cedex 5, France}
\address[4]{IRFU, CEA, Universit\'e Paris-Saclay, 91191 Gif-sur-Yvette, France}
\address[5]{Laboratoire Univers et Particules de Montpellier, Universit\'e de Montpellier, CNRS/IN2P3, CC 72,Place Eug\`ene Bataillon, F-34095 Montpellier Cedex 5, France}
\address[6]{Institut de Recherche en Astrophysique et Plan\'etologie, CNRS-INSU, Universit\'e Paul Sabatier, 9 avenue Colonel Roche, BP 44346, 31028 Toulouse Cedex 4, France}

%%%%%%%%%%%%%%%%%%%%%%%%%%%%%%%%%%%%%%%%%
\begin{abstract}

In this work, we describe the optical properties of the single photoelectron (SPE) calibration system designed for NectarCAM, a camera proposed for the  Medium Sized Telescopes (MST) of the Cherenkov Telescope Array (CTA). One of the goals of the SPE system, as integral part of the NectarCAM camera, consists in measuring with high accuracy the gain of its photo-detection chain. The SPE system is based on a white painted screen where light pulses are injected through a fishtail light guide from a dedicated flasher.  The screen -- placed 15\,mm away from the focal plane -- is mounted on an XY motorization that allows movements over all the camera plane. This allows in-situ measurements of the SPE spectra via a complete scan of the 1855 photo-multiplier tubes (PMTs) of NectarCAM.
This calibration process will enable the reduction of the systematic uncertainties on the energy reconstruction of $\gamma$-rays coming from distant astronomical sources and detected by CTA.

We discuss the design of the screen used in the calibration system and we present its optical performances in terms of light homogeneity and timing of the signal.

\end{abstract}


\end{frontmatter}

%% Start line numbering here if you want
%\begin{linenumbers}

%%%%%%%%%%%%%%%%%%%%%%%%%%%%%%%%%%%%%%%%%%%%%%%%%%%%%%%%%%%%%%%%%%
\section{Introduction}
\label{intro}

\begin{figure*}[t!]
\centering
\subcaptionbox{\label{screen_flasher}}{\includegraphics[scale=0.16]{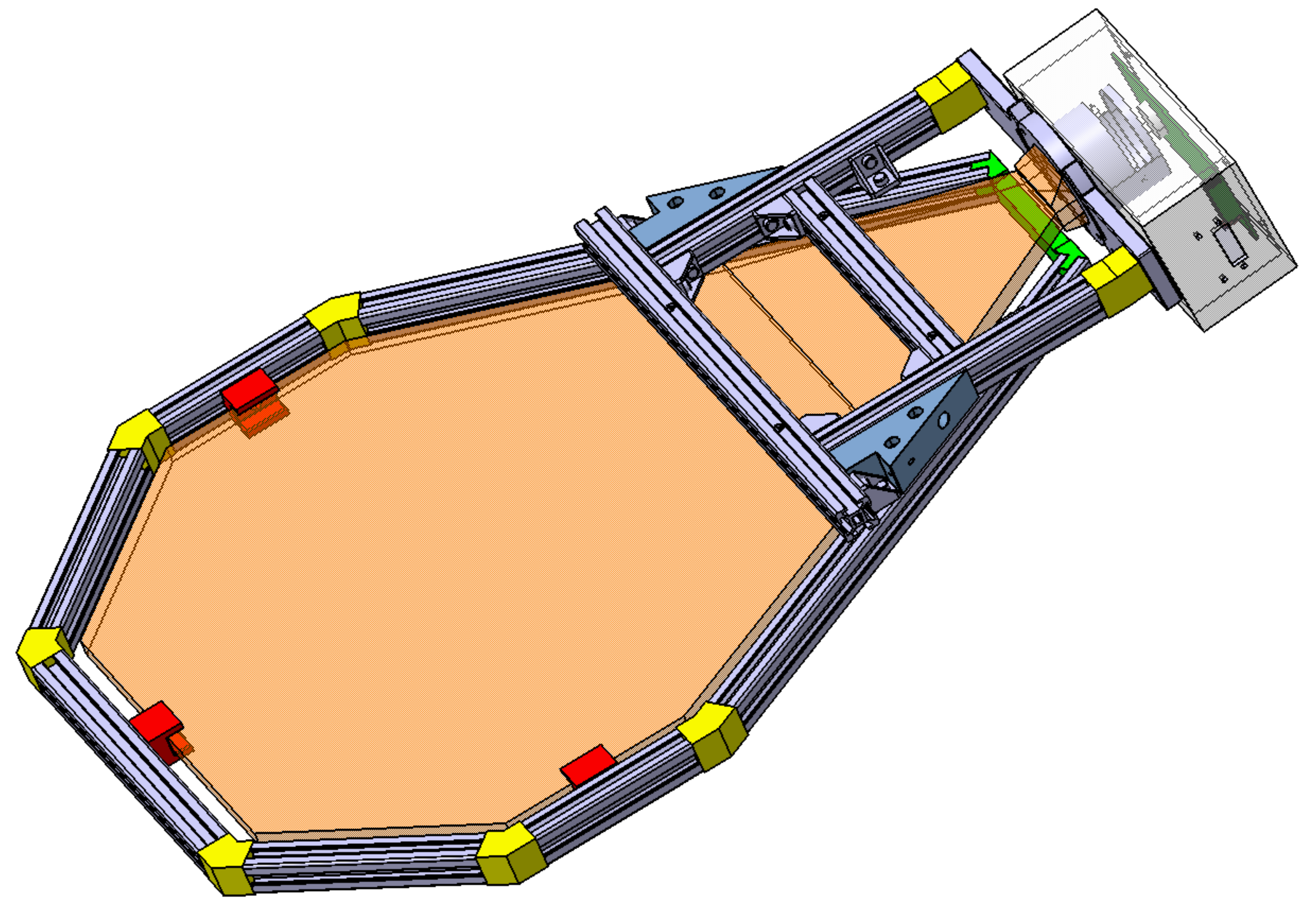}}
\hspace{5mm}
\subcaptionbox{\label{camera_front_view}}{\includegraphics[scale=0.23]{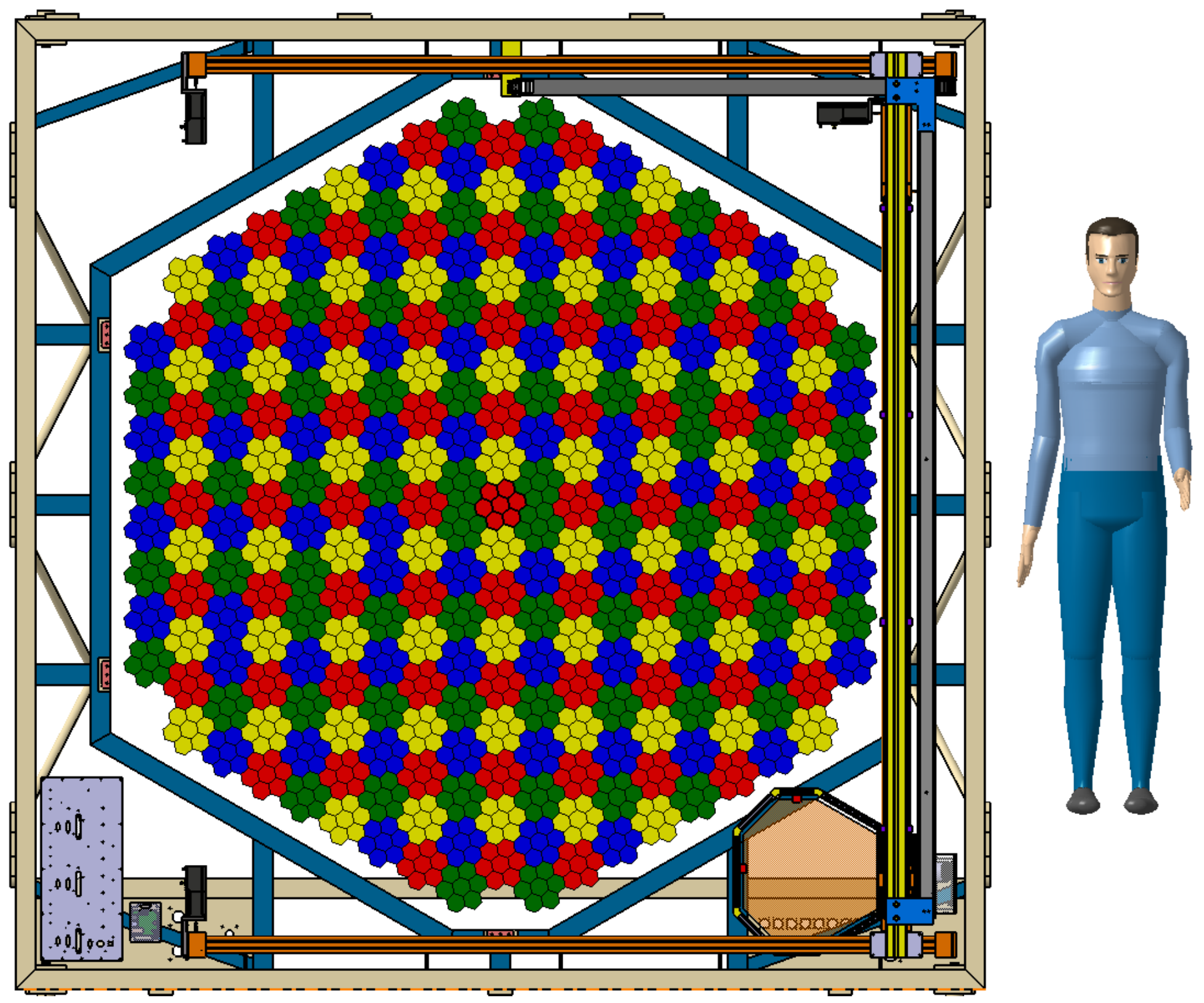}}
\hspace{5mm}
\subcaptionbox{\label{camera_lateral_view}}{\includegraphics[scale=0.18]{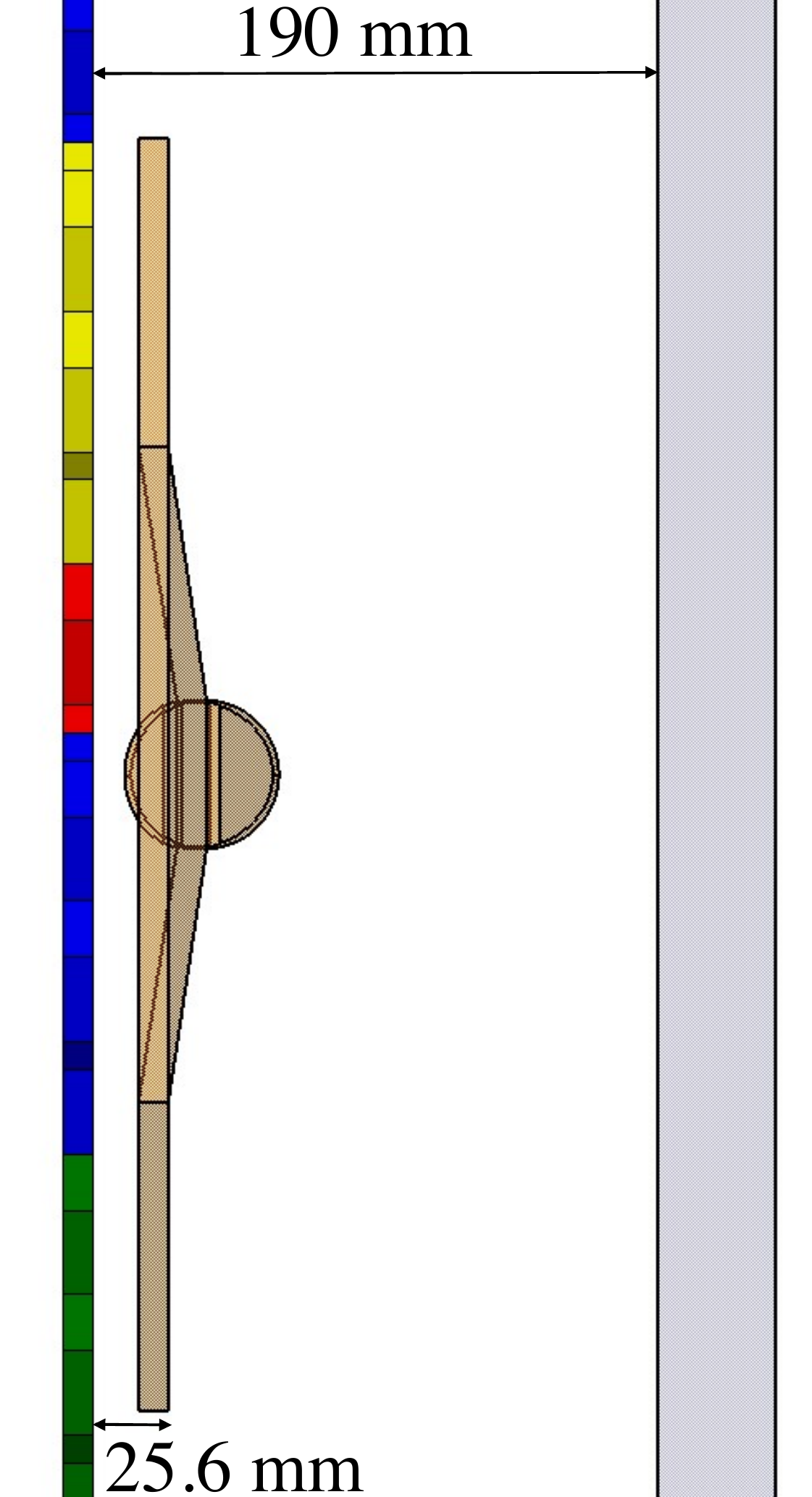}}
\caption{(a) bottom view of the assembly of the flasher box containing the LED electronic card, the fishtail light guide, and the final version of the screen; (b) front view of the camera focal plane with the SPE calibration system in parking position on the bottom-right corner (different colors are used to distinguish the PMT modules); (c) simplified lateral view of the front of the camera, showing the screen in a possible position during a camera scan: the emitting side is facing the PMTs, while the reflective one is looking at the mirror dish of the telescope. Two distances are reported: the distance between the entrance of the Winston cone and the edge of the screen facing the mirrors (25.6 mm), and the minimum distance between the camera window (gray band) and the screen (190 mm).}
\end{figure*}

The current most sensitive $\gamma$-ray telescopes employ the so-called imaging atmospheric Cherenkov technique that allows $\gamma$-rays to be indirectly detected through the light yield of their atmospheric showers \cite{Hinton2009}.
A $\gamma$-ray entering the atmosphere interacts with the Coulomb field of nuclei in the medium, and generates a shower of secondary particles. The velocity of $e^{-}$ and $e^{+}$ in the shower can be higher than the speed of light in the air, so they induce the emission of a Cherenkov-light cone that propagates towards the ground and can be observed with dedicated telescopes.
The currently operating ground-based $\gamma$-ray telescopes are VERITAS,\footnote{\url{https://veritas.sao.arizona.edu/}} H.E.S.S.,\footnote{\url{https://www.mpi-hd.mpg.de/hfm/HESS/}} MAGIC,\footnote{\url{https://magic.mpp.mpg.de/}} FACT,\footnote{\url{https://www.isdc.unige.ch/fact/}} and the future generation is represented by the Cherenkov Telescope Array (CTA) \citep{CTA2011, CTA2016}, which will enable observations in the interval between 30 GeV and 300 TeV with a sensitivity improved by a factor 5-20 depending on energy.
CTA will consist of two arrays with telescopes of three different sizes: the Large Sized Telescopes (LST) with a 23 m-diameter, the Medium Sized Telescopes (MST) with a 12 m-diameter, and the Small Sized Telescopes (SST) with a 4 m-diameter. In order to cover the whole sky, the two arrays will be placed in the Northern hemisphere in La Palma (Canary Islands, Spain) and in the Southern hemisphere, at Paranal (Chile), respectively.

The reconstruction of the $\gamma$-ray properties, such as arrival direction and energy, is not trivial, since the Cherenkov telescopes rely on an indirect detection technique. An important source of systematic uncertainty derives from the knowledge of the gain of the whole photo-detection chain, including detectors such as the photo-multiplier tubes (PMTs), which detect the Cherenkov light produced by the particle showers. 

The calibration process is crucial to improve the performances of current-generation Cherenkov telescopes and of the future CTA in the reconstruction of the $\gamma$-ray properties. In particular, various techniques are employed to monitor the gain of the PMTs.

VERITAS relies on a LED-based flasher system to measure the gain of the PMTs through both the photostatistics method and the single-photoelectron (SPE) spectrum fitting \cite{VERITAS2010}. In the first method, the gain is estimated by exploiting the relation between the mean number of photoelectrons hitting the first dynode and the width of the measured charge distribution \cite{Biller1995}, \cite{Mirzoyan1997}. In the second method, SPE measurements are performed in-situ by placing a thin aluminum plate with 3\,mm holes aligned with the center of each PMT.
The gain determination through the SPE fitting is also adopted by H.E.S.S. \citep{HES2004, HESS2017}. The system used at H.E.S.S. consists of a pulsing LED and a diffuser to homogeneously illuminate the camera, placed in a shelter 2 m away from it.
Finally, MAGIC uses strong light pulses, with three methods for the absolute light flux calibration \cite{MAGIC2002, MAGIC2003, MAGIC2005}. In the first method, the intensity of the pulse is obtained from the SPE spectrum of a so-called ``blind pixel'', darkened by a calibrated filter, while in the second one it is retrieved through a comparison with a calibrated PIN diode. Once the light intensity is known, the gain of all the PMTs can be retrieved. In the third method, the one currently in use, the gain of the PMTs can be determined through the photostatistics method, as done by VERITAS.

The multi-component nature and the large flexibility in operation of CTA will require different calibration systems, techniques, and strategies for the camera calibration, based on method of photostatistics, single photoelectron acquisitions, and observational data such as muon rings \citep{CTA2015, CTA2015b, CTA2017b}. For the MSTs of CTA, we developed an SPE calibration system that enables the measurement of the gain of the whole photo-detection chain with a statistical uncertainty lower than 1\%.
The SPE calibration system is integrated in the NectarCAM project \cite{Nectarcam2016}, one of the cameras proposed for the MSTs \cite{CTA2016}.\footnote{The other camera proposed for the modified Davies-Cotton MSTs is the FlashCam \cite{FlashCAM2012}.} NectarCAM has been designed to cover the energy range between 100 GeV and 30 TeV, with a wide field of view of $8^{\circ}$. NectarCAM encompasses 265 different modules: each of them includes PMTs, HV dividers, high voltage supply, pre-amplifier, trigger, readout, and Ethernet transceiver. Each module hosts 7 PMTs from Hamamatsu (R12992-100), with 7 dynodes each, for a total of 1855 PMTs in the whole camera. In front of each PMT, a Winston cone is mounted in order to maximize the light collected at the photocathode of each PMT.  

%An overview of the SPE calibration system, and the requirements for its reflective screen are presented in Sec. \ref{overview} and \ref{requirements}. The experimental setup is described in Sec. \ref{exp_setup}, and the research activity to achieve the final design of the diffuse-reflective screen is then discussed in Sec. \ref{ReD_activity}. In Sec. \ref{Optical_prop}, the optical properties of the final screen will be presented. Finally, a brief summary is provided in Sec. \ref{conclusions}.

%%%%%%%%%%%%%%%%%%%%%%%%%%%%%%%%%%%%%%%%%%%%%%%%%%
\section{Overview of the SPE calibration system}
\label{overview}

Both for the mirror alignment and the point spread function (PSF) studies, a movable system is required on the focal plane. The study of the gain of the PMTs can be done exploiting the same system. This brings another important advantage: the gain estimation can be performed with the camera closed (including during daytime, provided a lightproof camera enclosure), without contamination from the night sky background.
The dual purpose of the SPE calibration is accomplished by the two sides of its screen: the side directed towards the PMTs will enable the measurement of the gain of the whole photo-detection chain in SPE regime, while the other side will be used for mirror alignment and the study of the optical PSF.\footnote{The optical PSF refers to the instrumental response to a distant point-like source emitting in the visible spectrum, and for Cherenkov telescopes it is of the order of 1 arcmin. The $\gamma$-PSF refers to the accuracy in the reconstruction of the $\gamma$-ray direction, and it is of the order of 0.05-0.1 deg.}
%The latter is usually defined as the standard deviation of a  2-dimensional Gaussian fitted to the distribution of the reconstructed event directions of the $\gamma$-ray excess (see for example \cite{MAGIC2016}).}
The developed system is formed by two main components: (i) a source light box that produces flashes of light, at a typical wavelength of 390\,nm, which are injected into (ii) a 10\,mm thick screen made of Poly(methyl methacrylate) (PMMA) covering an area equivalent to 51 PMTs, or $\sim$1.7\,deg$^2$ (Fig.~\ref{screen_flasher}).
The system is mounted inside the camera ($15$ mm away from the Winston cones) on a system of $xy$ rails, and moved by motors over the entire focal plane. When not in use, it is located in a parking position, on the bottom-right corner of the camera, as shown in Fig. \ref{camera_front_view}. Figure \ref{camera_lateral_view} shows the position of the screen in a lateral view of the camera.

Both sides of the screen are painted with a highly reflective paint. The side facing the focal plane is painted with a dedicated pattern to optimize the homogeneity of light emission, while the side facing the mirrors is covered with three layers of paint. 

Both the mirror alignment and the PSF studies can be performed by analyzing the image of a point-like light source that forms on the side of the screen facing the mirrors. A movable system enables the study of the PSF both on- and off-axis. The point-like source image is captured by a charge-coupled device camera located in the center of the mirror dish, and further analyzed.
The mirror alignment can be performed by using different methods, such as the 2f alignment \cite{Anderhub2013}, the SCCAN method \cite{Arqueros2005} that tracks a star and observes the reflections of the mirrors on a camera placed in the reflector’s focal point, or the Bokeh method \cite{Ahnen2016} that analyzes the blurring of an image on the focal plane. Finally, as in the MST case, the alignment methods use remotely controllable orientation actuators on the mirrors in order to optimize the live image of a star on the dedicated screen \cite{Biland2008}.

%is usually performed by studying  For example, in the so-called Bokeh method \cite{Ahnen2016} the blurring of an image out of focus is described by a Bokeh function B of the imaging system. All the mirrors of the telescope can be aligned by matching the reflector’s actual Bokeh B to an ideal Bokeh template, which is the Bokeh of the well aligned reflector.}

The emission towards the focal plane, needed for the gain estimation, is obtained through the direct injection of light into the screen. For this aim, a dedicated light box source  has been designed at the {\it Laboratoire Univers et Particules de Montpellier} (LUPM). It is equipped with 12 light emitting diodes (LEDs)\footnote{Standard 3\,mm UV LEDs (ref. 3RS4VCS): wavelength 390\,nm, typical forward voltage 3-4\,V, opening angle 30$^\circ$, light intensity 0-200\,mCd, driven voltage 7.5-16\,V.} at 390\,nm pulsing with adjustable frequency and amplitude. The typical full width at half maximum (FWHM) of the pulses is between 3 and 5 ns.
The light produced by the flasher is conveyed and injected into the screen through a fishtail light guide, made of the same PMMA material as the screen.
To save as much space as possible in favor of a larger screen-surface, the width and the length of the fishtail light guide are set to equal values of 148.1 mm. The input diameter of the fishtail is equal to 50\,mm.

To achieve a low-intensity emission towards the focal plane, the screen is entirely covered with a reflective paint that forces the light to be reflected several times inside the screen, so that only a small fraction can escape from its surface. With this design, low-intensity flashes illuminate the PMTs so that the gain can be measured acquiring SPE spectra. To facilitate the light injection, the fishtail light guide is painted with the same material as the screen, and then wrapped with a black tape to avoid light leakage. 

%%%%%%%%%%%%%%%%%%%%%%%%%%%%%%%%%%%%%%%%%%%%%%%%%%%%%%
\section{Requirements for the diffuse-reflective screen}
\label{requirements}

In the conceptual design of the diffusive-reflective screen, we needed to account for three main requirements from the project:

\begin{itemize}[noitemsep]
\item [-] the dimension of the screen must be greater than a disk with a diameter of 170\,mm (corresponding to 0.6\,deg projected on the sky) to perform correctly the mirror alignment and the study of the optical PSF;
\item [-] for the PSF studies, the screen surface has to provide a Lambertian diffusive reflectivity $>90$\% between 450 and 700 nm.
\item [-] the distance between the focal plane and the reflective side of the screen (including the thickness of the screen) must be 25.6 mm. This extra distance is needed to place the screen in the image plane of distant stars, since the front of the Winston cone is focused at $\sim$10\,km, the typical altitude of the maximum development of atmospheric air-showers.
\end{itemize}

\noindent We aimed to build a screen with as homogeneous an emission as possible to enable SPE acquisition over all its surface.
We investigated systematically the light-intensity contour maps to evaluate the amount of area whose emission is within a certain fraction of the maximum value measured on the screen.

Starting from these criteria, the design process focused on the following aspects:

\begin{itemize}[noitemsep]
\item [-] geometry: whatever the shape of the screen, it should be significantly larger than the PSF. It should be as large as possible to enable a fast scan of the camera, and it must not shadow the focal plane once in parking position.
\item [-] coating type: it should be characterized by a high reflectivity both to ensure a low-intensity emission towards the focal plane, and to perform the mirror alignment on the other side;
\item [-] coating pattern: a study of the coating pattern has been performed in order to achieve an emission as homogeneous as possible; 
\item [-] coating application: different types of coating-application methods have been investigated, not only to improve the light homogeneity, but also to ensure high reproducibility and stable optical performances.
\end{itemize}

\noindent All these aspects have not been investigated separately, but they entered in different moments of the R\&D phase of the reflective screen design, as explained in the following sections. 

%%%%%%%%%%%%%%%%%%%%%%%%%%%%%%%%%%
\section{Experimental setup}
\label{exp_setup}

In the R\&D phase of the SPE calibration system, we tested several screen prototypes, different in shape, painting-type coverage, painting pattern, and painting application method. In particular, as described in the following sections, we investigated four geometries (square, rectangular, circular, and octagonal), two reflective materials (polytetrafluoroethylene and reflective painting), three coating methods (brush, air brush, and dip-coating), and dedicated painting patterns for different prototypes.

\begin{figure}[t!]
\centering
\includegraphics[scale=0.25]{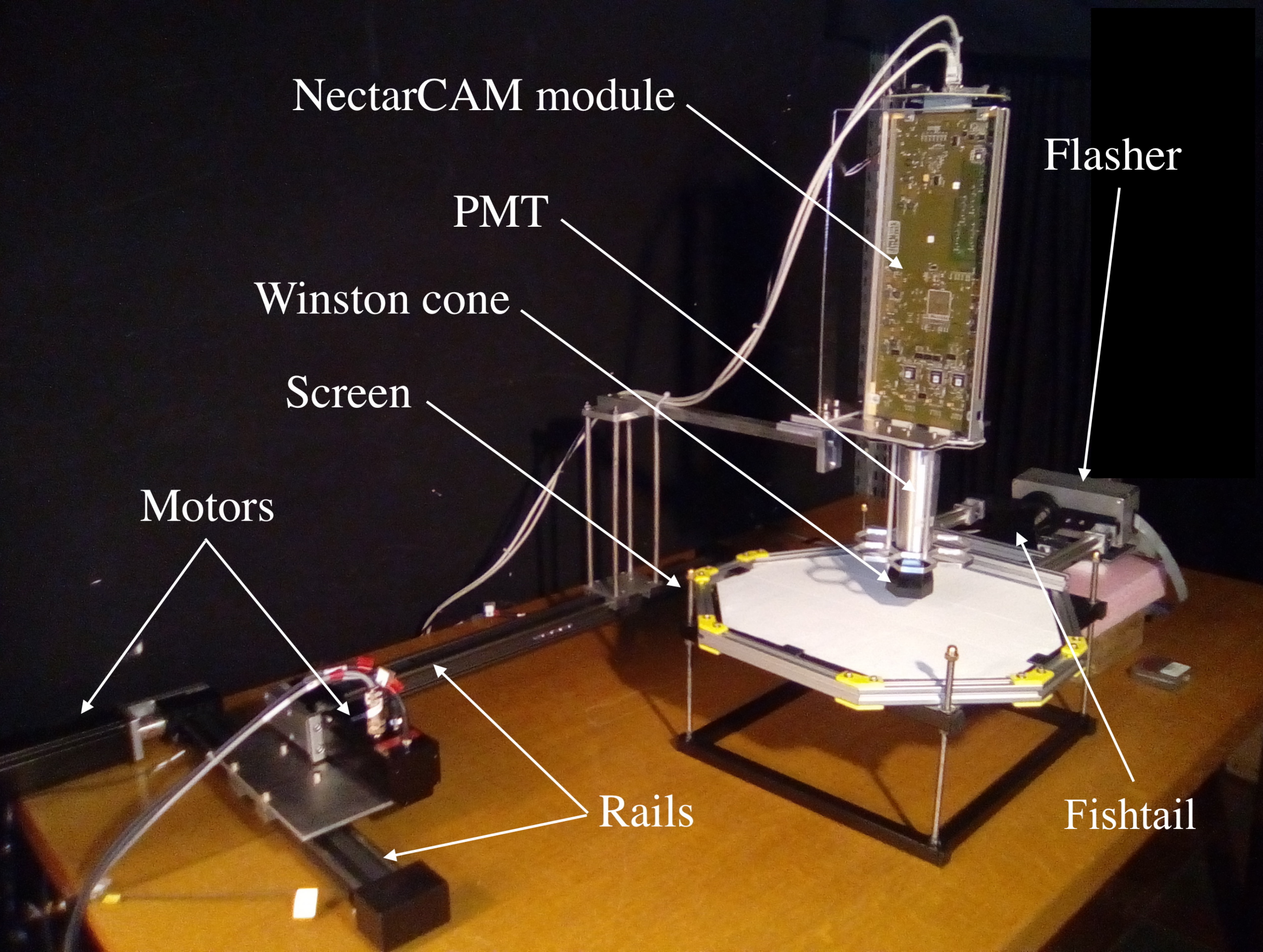}
\caption{Test-bench equipment placed in the dark room. A system of two motors moves a NectarCAM front-end board, equipped with only one PMT over the screen. Light pulses are injected from the flasher into the fishtail light guide (black piece) and, in turn, into the screen. The LED electronic card of the flasher is stored inside the aluminum box attached to the fishtail.}
\label{test_bench}
\end{figure}

%For the tests, the screens were all placed in a dark room, and only one PMT, placed at a fixed distance of 15 mm from the screen surface, was used to acquire light intensity measurements. The PMT (Hamamatsu R12992-100) is mounted on a system of {\it xy} rails (model ZLW-1040-S with 100 mm-long trolley), with two motors controlled via a dedicated software. This motorized system allows us to move the PMT all around the screen for a complete scan. A picture of the test bench is shown in Fig. \ref{test_bench}.
For the tests, the screens were all placed in a dark room. A NectarCAM front-end board \cite{Naumann2012} (FEB, originally developed by a consortium involving {\it Institut de recherche sur les lois fondamentales de l'Univers} \mbox{-- IRFU --}, {\it Laboratoire de physique nucl\'eaire et de hautes \'energies} \mbox{-- LPNHE --}, {\it Instituto de Ciencias del Cosmos, Universitat de Barcelona} \mbox{-- ICCUB --}, and LUPM), is equipped with one PMT (Hamamatsu R12992-100) placed at a fixed distance of 15 mm from the screen surface.
%For the first set of measurements on smaller prototypes, the PMT was moved by motors on a 20 cm $\times$ 20 cm system of rails. The movements were controlled by a fast prototyping framework, called Pyrame \cite{Pyrame}, developed at {\it Laboratoire Leprince-Ringuet} (LLR). As we moved to larger prototypes, the rails have been substituted by longer ones, $60$ cm $\times$ 40 cm, and the motors have been replaced by drylin\textsuperscript\textregistered stepper motor NEMA 23XL, the same that will be used on the NectarCAM, which guarantee a positioning accuracy of $\pm 0.035$ mm.
\noindent They were moved by motors (drylin\textsuperscript\textregistered stepper motor NEMA 23XL) on a 60 cm $\times$ 40 cm system of rails (model ZLW-1040-S with 100 mm-long trolley) all around the screen to acquire light intensity measurements. Motors are controlled via a dedicated software and guarantee a positioning accuracy of $\pm 0.035$ mm. A picture of the test bench is shown in Fig. \ref{test_bench}.

\noindent In order to ensure a positioning reproducibility during the test bench activities, a dedicated mask was built for each prototype. A grid drawn on such masks provided a way to cross-reference the motors and the screen frame. We estimate the accuracy of the screen positioning of $\pm 2$ mm, both along the $x$ and $y$ axes.

The light pulses injected into the screen are produced by a flasher developed at LUPM. The light is produced by 12 LEDs emitting at 390\,nm, the typical wavelength of the Cherenkov light. The intensity can be tuned changing the operating voltage of the LEDs between 7.5 and 16.5 V. An extra LED\footnote{Kingbright LED (ref. 934LSRD): wavelength 660\,nm, typical forward voltage 1.85\,V, opening angle 60$^\circ$, light intensity 0-20 mCd at 2\,mA, driven in continuous current mode 0-12.5\,mA.} in DC mode at longer wavelength ($\sim$660\,nm) has been added to simulate the night sky background \cite{Bouvier2013}. The intensity of the light pulses (both in terms of number of LEDs, and their voltage) can be adjusted through a labview software interface or an OPC UA server. The typical pulse frequency was set to 100 Hz, and the trigger pulse duration to 250 ns.

\noindent Several versions of the flasher have been produced. Depending on the version of the test-bench setup, different neutral density filters (THORLABS filters with optical density, OD, 0.5, 1.0, 2.0, 3.0, respectively) have been used to achieve the optimal emission. The final version of the flasher includes a filter with OD=1 that allows us to achieve an emission ranging from $\sim$0.01 photoelectrons (1 LED at minimum voltage) up to $\sim$50 photoelectrons (12 LEDs at maximum voltage). By removing the filter, the intensity can be increased up to $\sim$500 photoelectrons.

For each set of measurements, we adjusted the intensity of the light pulses depending on the PMT voltage (typically between 1000 and 1400 V) with an appropriate intensity of the LEDs. The nominal high-voltage for the NectarCAM PMTs in operating conditions is around 1000 V. With an appropriate intensity of the LEDs, we could acquire spectra both in the faintest and in the brightest regions of the screen.

\begin{figure}[t!]
\centering
\hspace{-2mm}
\includegraphics[scale=0.28]{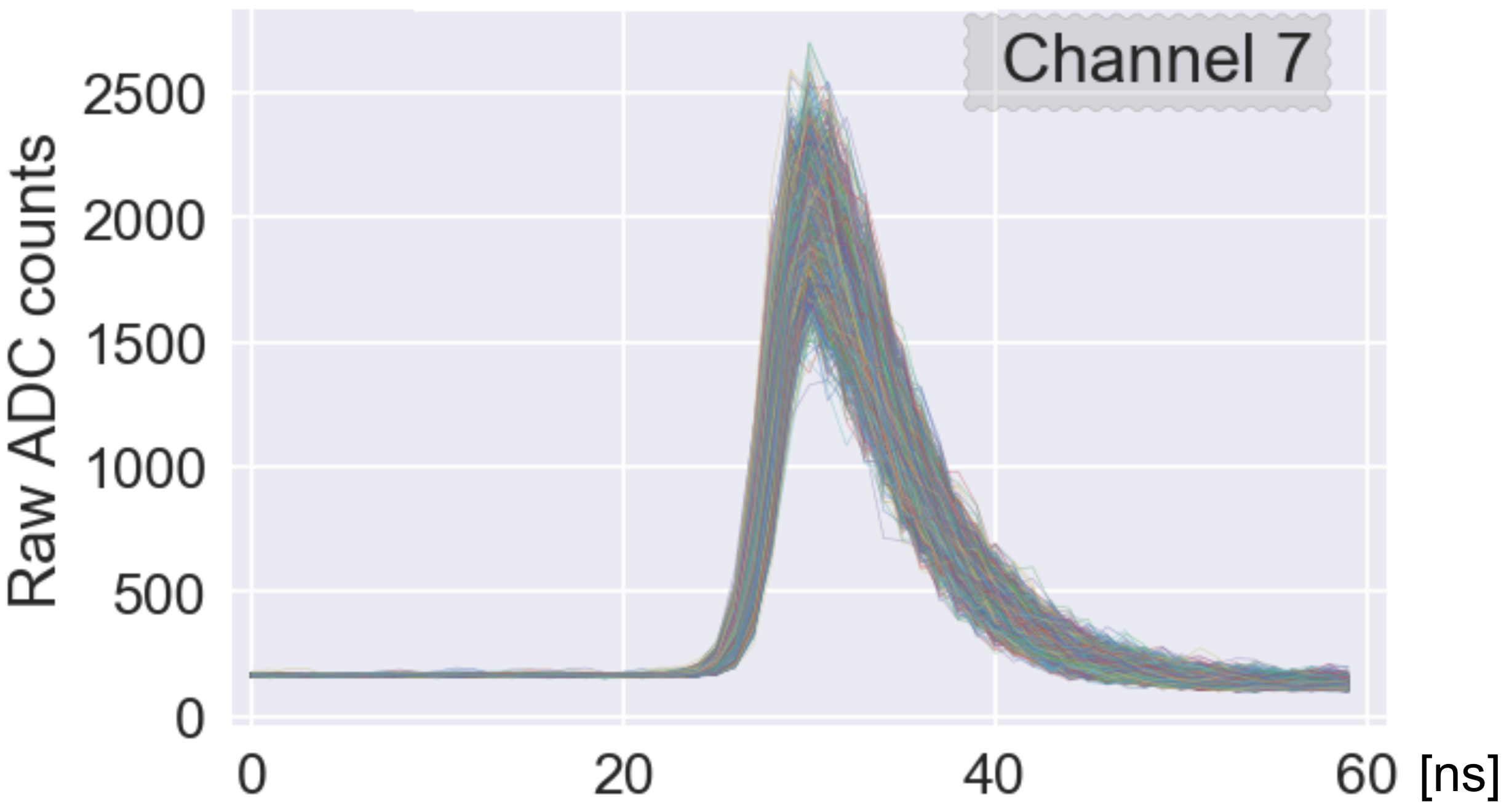}
\caption{Example of PMT traces induced by the flasher. The traces were collected in a single position over the screen, and the signal was integrated over a 60 ns window.}
\label{traces}
\end{figure}

Light intensity measurements were performed in high intensity regime, typically $> 100$ photoelectrons produced at the photocatode. Measurements were taken on a grid of points (usually $4$ cm $\times$ 4 cm), to obtain light intensity profiles or maps. For each position on the grid, we obtained the histogram of the total charge integrating a large number of collected signals (typically 30,000) over a fixed gate duration (typically between 60 and 100 ns). The mean value of the acquired gaussian distribution was taken as the indicator of the light intensity. The associated statistical uncertainty is less than 1\%, while the systematic one (due to the accuracy in the positioning, and to the setting of the acquisition windows) is better than 2\%. Figure \ref{traces} shows the traces collected over a single position of the screen.

%%%%%%%%%%%%%%%%%%%%%%%%%%%%%%%%%%%%%%%%%%%%%%%%%%%%%%%%%%
\section{Results}
\label{ReD_activity}

\subsection{Square screen}
\label{square}

At first, a small $20$ cm $\times$ 20 cm reflective screen prototype was built to validate the concept design. Such a surface enables the calibration of one entire module of the camera, or 11 individual PMTs.
Different diffusive materials have been tested with this first prototype. In particular, we used a $\sim$200\,$\mu$m polytetrafluoroethylene foil and a Bicron paint (Saint Gobain BC-620) applied with a brush. This paint was selected for its high Lambertian diffusive reflectivity ($> 90$\% above 400 nm) \cite{Ghal2006}, its cost, and its ease of use. We also used this paint to paint the fishtail light guide; it was then covered with black tape\footnote{The performances of the black tape as a function of the aging and different environmental conditions will be explored for in-situ operations and could be replaced by black paint if need be.} to avoid light leakage.
Furthermore, several combinations of top-bottom and edge coverage of the screen have been explored:

\begin{enumerate}[noitemsep]
\item polytetrafluoroethylene on the bottom, black tape on the edges (PTFE $+$ BT); %nothing on the top
\item paint on the bottom, black tape on the edges (PAINT $+$ BT); 
\item paint + aluminum foil (PAINT $+$ AF);
\item paint on the bottom, paint on the edges (PAINT $+$ PAINT); %nothing on the top
\item paint on the bottom, polytetrafluoroethylene on the top, paint on the edges (PAINT $+$ PAINT $+$ PTFE);
\item paint on the bottom, paint on the top, paint on the edges (PAINT $+$ PAINT $+$ PAINT).
\end{enumerate}

\begin{figure}[t!]
\centering
\hspace{-4mm}
\includegraphics[scale=0.25]{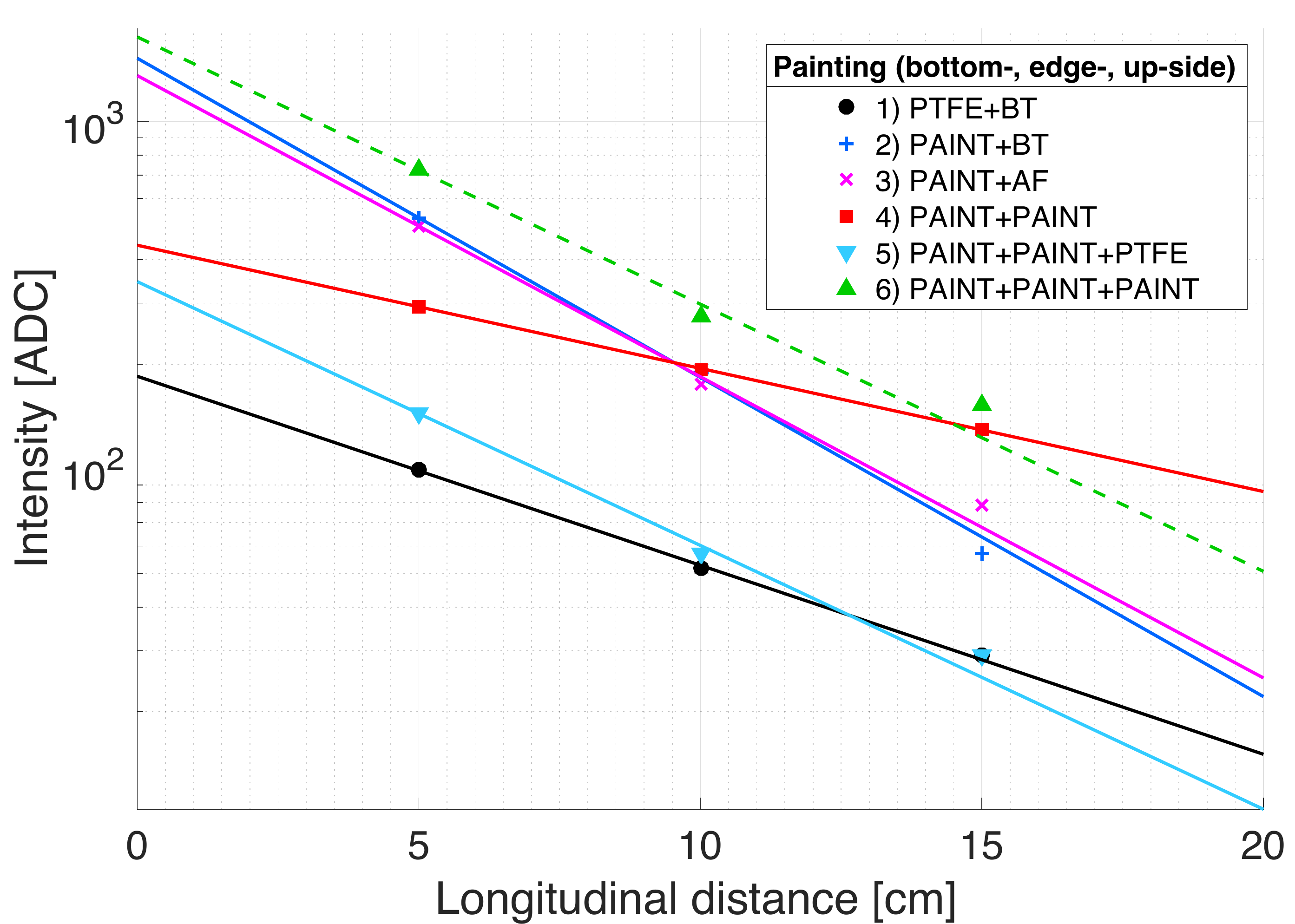}
\caption{Light-intensity profiles measured along the central axis of the square screen. Measurements were taken at a distance of 5, 10, and 15 cm from the injection edge, and are expressed in analog-to-digital counts (ADC). Measurements were taken switching on 1 LED at 8 V, with a filter of OD=0.5, and setting the PMT high-voltage at 1000 V (solid line). For configuration 6, measurements were taken in higher luminosity conditions: 1 LED at 10.3 V, without OD filter, and PMT high-voltage set at 1200 V (green dotted line).}
\label{square_profile}
\end{figure}

\noindent Figure \ref{square_profile} shows the light-intensity profiles measured along the central axis, perpendicular to the light injection edge, for the above-mentioned configurations. For each configuration, the attenuation length was estimated through a linear regression, and results are reported in Table~\ref{att_lenghts_square}. The estimated uncertainty on the slope is smaller than 0.02\,cm for all measurements.

Among the first four configurations that cover only the bottom and the edges, those involving PTFE $+$ BT and  PAINT $+$ PAINT are able to minimize the emission decrease with increasing longitudinal distance. To better control the emission (in order to reach the SPE regime), and ensure that the light would be carried to longer distances in larger screens, we observed that a coverage on the top side was also necessary (configurations 5 and 6). 
Both configurations provide a similar attenuation length, and capability to spread the light.
However, after the first uses of these materials it became evident that it was quite difficult to properly glue a polytetrafluoroethylene foil over a large surface. For this reason, after this first attempt, we focussed our attention only on the Bicron paint.

\begin{table}[]
\center
\begin{tabular}{|c|c|}
\hline
Configuration & \begin{tabular}[c]{@{}c@{}}Attenuation\\ length {[}cm{]}\end{tabular}\\
\hline
1             & 13.23   \\ \hline
2             & 5.47     \\ \hline
3             & 6.40      \\ \hline
4             & 14.43    \\ \hline
5             & 9.48       \\ \hline
6             & 7.08       \\ \hline
\end{tabular}
\caption{Attenuation length for each configuration tested in the square screen prototype. See text for details.}
\label{att_lenghts_square}
\end{table}

%===============================
\subsection{Rectangular screen}
\label{rectangular}

After preliminary tests on the small prototype described in Sec. \ref{square}, a larger one - $60$ cm $\times$ 20 cm - has been produced to allow a larger number of PMT modules to be scanned at the same time. The light is injected with the same system as that used for the square screen, i.e., the fishtail light guide, covered with the same reflective paint used for the screen and wrapped with black tape. This geometry enables to scan either 3 PMT modules, or 43 individual PMTs per scan.

\begin{figure}[t!]
\center
\includegraphics[width=1.00\columnwidth]{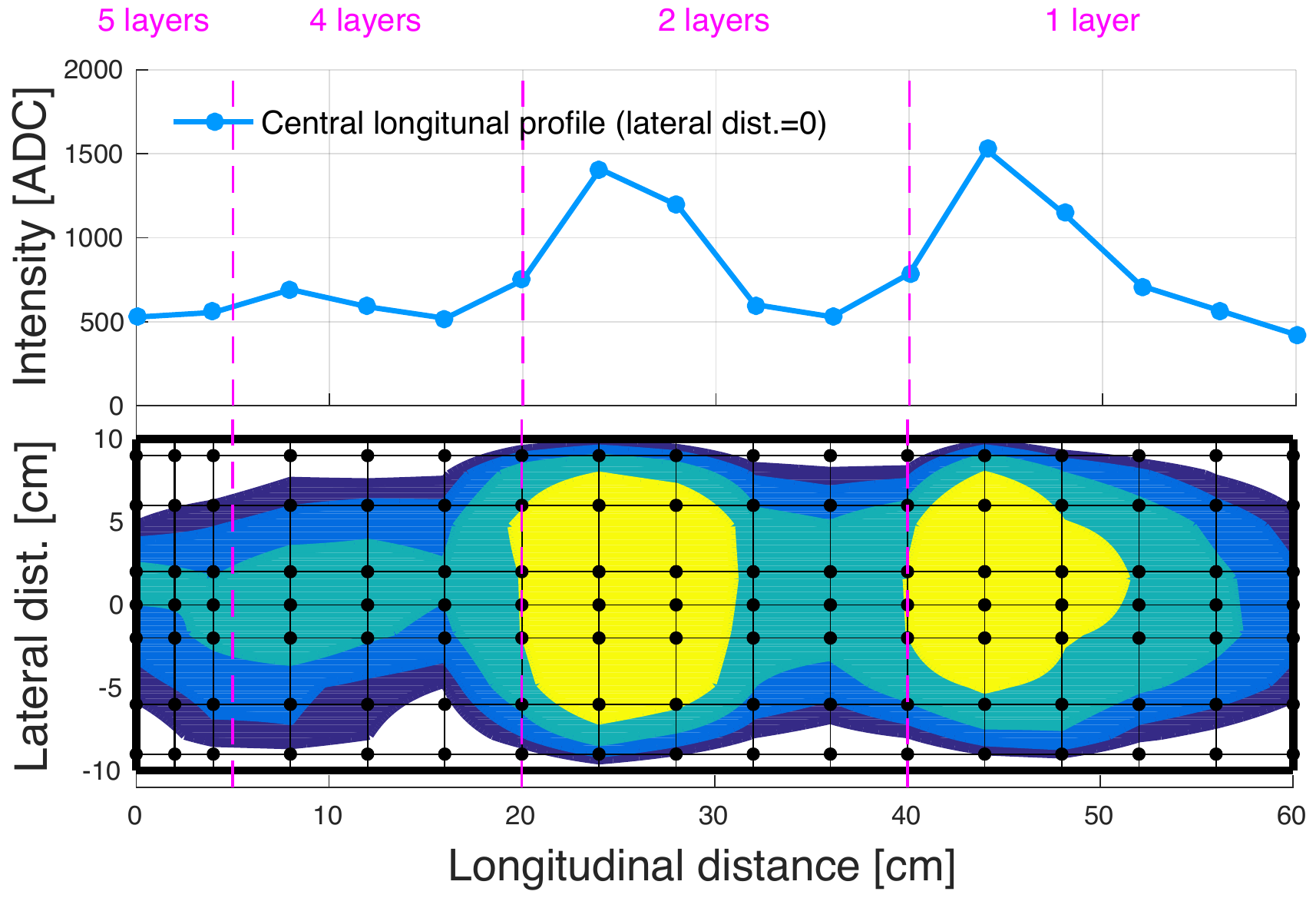}
\caption{Longitudinal light-intensity profile (top) and light intensity contour-map (bottom) for configuration (iii) of the rectangular screen (color online). Contours enclose the regions where the light intensity is within a given percentage of the maximum intensity, $I_{\rm max}$, measured on the screen: $I>50\% $ (yellow), $33\% <I<50\%$ (green), $25\% <I<33\%$ (light blue), $20\% <I<25\%$ (blue), $I<20\%$ (white). Light is injected into the left edge of the screen \mbox{(longitudinal distance} =0). Measurements (black dots) were taken on a grid 4 cm $\times$ 4 cm, with some extra measurements close to the injection edge, and along the central axis. Magenta dotted lines refer to the painting pattern.}
\label{rectangular_contour_maps}
\end{figure}

In accordance with the results obtained with the small square prototype, we decided to use a specific painting pattern. In fact, the short attenuation length characterizing the configuration PAINT $+$ PAINT $+$ PAINT, points out that a layer of paint on top of the screen is not sufficient to propagate the light up to the extreme edge of a longer screen. The adopted strategy consisted then in brush-painting the screen with a decreasing number of superimposed layers of paint as a function of the distance from the injection edge. In this way, the coating is thicker close to entrance, forcing the light to be carried further into the screen. In order to achieve an emission as homogeneous as possible, three different patterns have been tested:

\begin{itemize}[noitemsep]
\item  [(i)] 3 layers between 0 and 20 cm, 2 layers between 20 and 40 cm, and 1 layer between 40 and 60 cm (configuration 3-2-1);
\item  [(ii)] 4 layers between 0 and 20 cm, 2 layers between 20 and 40 cm, and 1 layer between 40 and 60 cm (configuration 4-2-1);
\item  [(iii)] 5 layers between 0 and 5 cm, 4 layers between 5 and 20 cm, 2 layers between 20 and 40 cm, and 1 layer between 40 and 60 cm (configuration 5-4-2-1),
\end{itemize}

\noindent where the origin is set at the light injection edge. Figure \ref{rectangular_contour_maps} shows the central longitudinal light-intensity profile, and the contour maps obtained for configurations (iii). The area covered within a factor $1/2$, $1/3$, $1/4$, $1/5$ from the maximum light intensity measured on the screen is 22.4\%, 55.3\%, 75.9\%, and 85.6\%, respectively.

%=============================
\subsection{Circular screen}
\label{circular}

Although the results obtained with the rectangular screen were satisfying, exchanges with the LST camera team pushed us to decrease the ratio between the length and the width of the screen, and to enlarge its surface to probe the tails of the optical PSF. Both the area and the shape of the previous prototype were not sufficient to guarantee the probing of the PSF tails, so a 38.4 cm-diameter circular screen, was built. This screen can calibrate 4 modules, or 41 individual PMTs at the same time. In this configuration, inspired from that proposed for one of the cameras of the SSTs \citep{ASTRI2013, ASTRI2017}, the light injection occurs all along the edge. The fishtail light guide was replaced by two Saint-Gobain scintillating optical fibers of diameter $\phi$ 1.2 mm, coming out from the light box, and running in opposite directions along the edge of the screen. The fibers were placed inside two rails grooved at 1/3 and 2/3 of the screen thickness, respectively. Black tape was used to cover the edge of the screen, and thus to avoid light leakage, as well as to keep the fibers in place inside the grooves. The part of the fibers between the light source and the edge of the screen was inserted into a black cladding.

%\begin{figure}[t!]
%\centering
%\includegraphics[scale=0.45]{Circ_1Layer_contour_map.pdf}
%\caption{intensity contour map of the 1 layer-painted circular screen. The color code is the same as Fig. \ref{rectangular_contour_maps}. Measurements (black dots) were taken on a grid of 4 cm $\times$ 4 cm.}
%\label{whole_circ_1lay.pdf}
%\end{figure}

\begin{figure}[t!]
\centering
\includegraphics[scale=0.4]{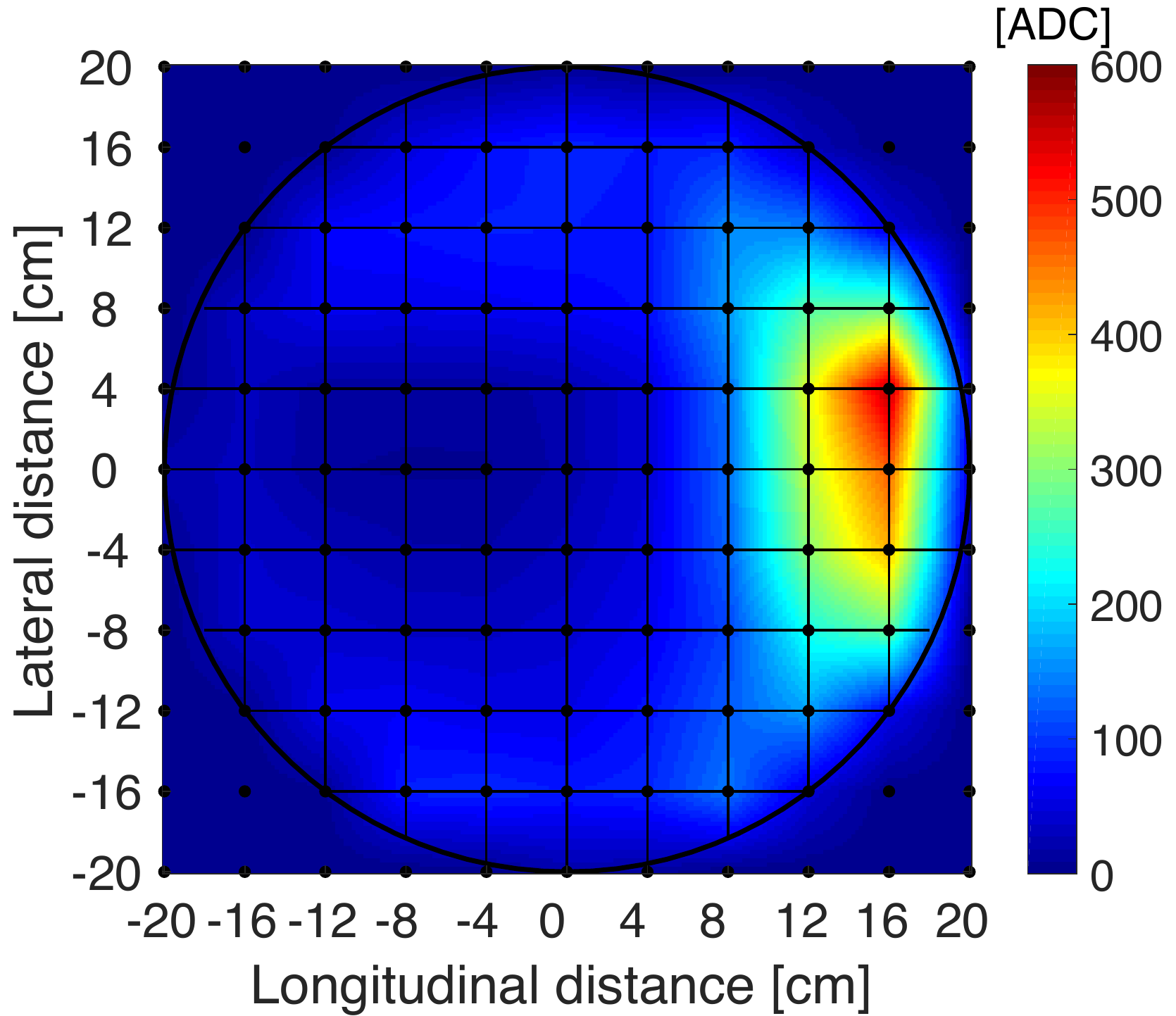}
\caption{Light intensity map (color online) of the 1 layer-painted circular screen. Measurements (black dots) were taken on a grid of 4 cm $\times$ 4 cm.}
\label{whole_circ_1lay.pdf}
\end{figure}

\begin{figure*}[t!]
\center
\subcaptionbox{\label{1_layers}}{\includegraphics[width=0.65\columnwidth]{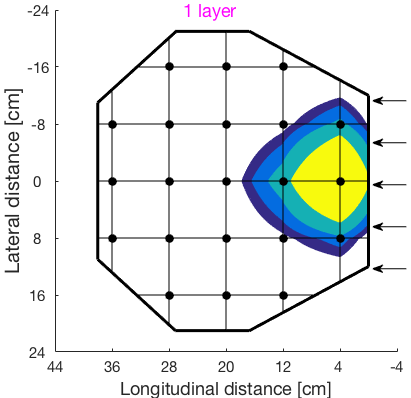}}
\hspace{2mm}
\subcaptionbox{\label{2_1_layer}}{\includegraphics[width=0.65\columnwidth]{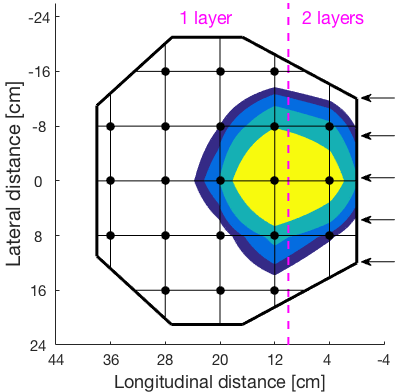}}
\hspace{3mm}
\subcaptionbox{\label{3_1_layer}}{\includegraphics[width=0.65\columnwidth]{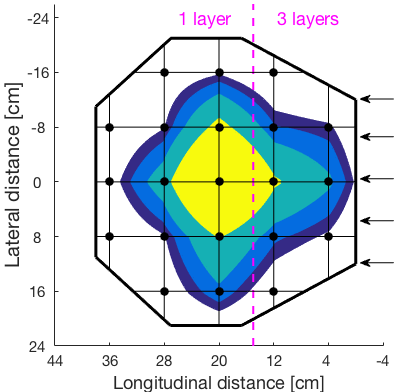}}
\hfill
\subcaptionbox{\label{3_2_1_layer}}{\includegraphics[width=0.65\columnwidth]{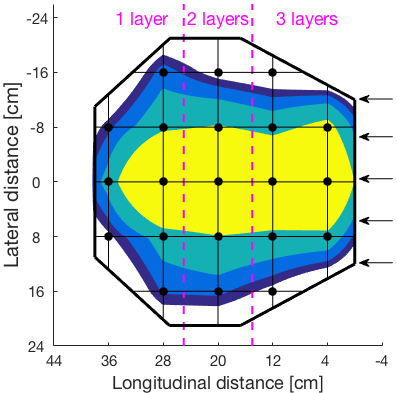}}
\hspace{5mm}
\subcaptionbox{\label{3_2_1_ex_layer}}{
  \hbox{
        \includegraphics[valign=t,width=0.65\columnwidth]{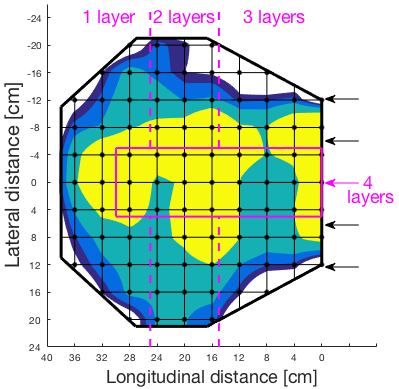}
        \hspace{3mm}
        \includegraphics[valign=t,width=0.4\columnwidth]{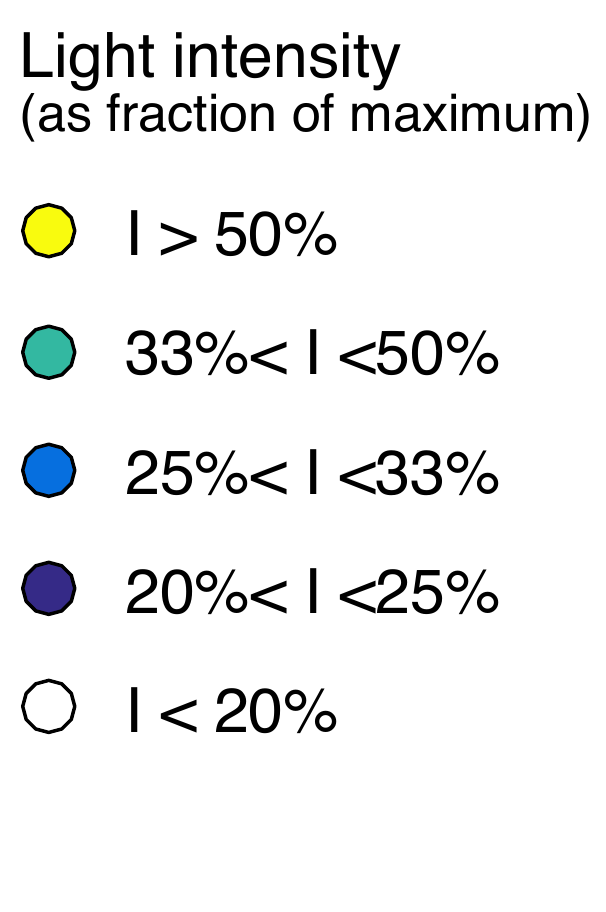}
       }
}
\caption{Light-intensity contour maps for different painting patterns of the octagonal screen (color online). The color code is the same as in Fig. \ref{rectangular_contour_maps}. Magenta dotted lines refer to the painting pattern (see also text for details). Black arrows indicate the light-injection edge. Measurements were taken in correspondence with the black dots.}
\label{pattern_phases}
\end{figure*}

The emission measurements revealed that this configuration fails to carry a significant amount of light towards the center of the screen and results in a light excess close to the initial contact point between the fibers and the screen. Several possible causes for such an excess have been identified: (i) the black cladding covering the fibers is not monolitic, hence some light could escape from the juncture; (ii) the holes from which the fibers exit the light box could be a bit larger than the fibers diameter; (iii) the black tape at the contact point between the fibers and the screen does not guarantee a perfect adherence, leaving a small part of the fibers uncovered. 
In order to stem the diffuse light coming from (i), the region between the light box and the screen - {\it i.e.}, where the optical fibers run - was wrapped with a foil of tedlar, a completely opaque material. The possible light dispersion due to (ii) was mitigated by adding a 3 cm-layer of black silicon at the exit of the fibers from the light box. The issue (iii) was harder to treat, and at the end a satisfying and reliable solution has not been achieved.
Despite these technical measures, the light excess was partially reduced, but not definitively cut out.
The light intensity map obtained for the entire 1 layer-painted screen is shown in Fig. \ref{whole_circ_1lay.pdf}.
%Beyond the problem due to the light leakage, the map points out that in this system the light is not carried out homogeneously along the fibers.
The light intensity becomes increasingly lower as we move away both from the screen edges and the fiber entrance. This results in a big dark region located in the central-left zone of the screen. 

\noindent While the observed light distribution inhomogeneity might be reduced with a more complicated paint pattern, the low control on the light leakage together with an increased complexity of the system due to extra hardware and materials ({\it i.e.}, optical fibers, tedlar, and black silicon) increases the breakdown risk during operating condition, and complicates the reproducibility of the system. For these reasons, we decided to explore what became our final design.

%===============================
\subsection{Octagonal screen}
\label{octagonal}

The aim of the following step in the design process was the development of a new screen embedding both the dimension of the circular screen and the reliability and the performance of the rectangular one. An octagonal screen whose length and width measure 40 and 42 cm, respectively, for a total area of 1338 cm$^2$, has been built. The area of this screen is large enough to contain 7 entire PMT modules, or 51 individual PMTs. The light-injection system is the same as for the rectangular one ({\it i.e.}, the fishtail light guide). The screen side facing the mirrors has been painted with 3 layers of reflective paint. The edges have been painted with the same reflective paint, with the exception of the three farthest edges - i.e., opposite to the light entrance - to avoid backward light-reflections. Finally, all the edges are covered with black tape.

Given the different geometry with respect to the rectangular system, the coating pattern (brush painted) towards the focal plane has been investigated step by step. Starting from one homogeneous layer, at each step, we added strips of coating whose position and width were determined by analyzing the light profiles along the central longitudinal-axis, and the lateral one if necessary.

\begin{itemize}[noitemsep]
\item [a)] 1 homogeneous layer over the entire screen surface;
\item [b)] 2 layers in the first 10 cm, 1 layer on the rest of the screen;
\item [c)] 3 layers in the first 15 cm, 1 layer on the rest of the screen;
\item [d)] 3 layers in the first 15 cm, 2 layers between 15 and 25 cm, 1 layer on the rest of the screen;
\item [e)] 3 layers in the first 15 cm, 2 layers between 15 and 25 cm, 1 layer on the rest of the screen, plus an additional layer between 0 and 30 cm along the longitudinal distance, and between -5 and 5 cm along the lateral distance.
\end{itemize}

\noindent Figure \ref{pattern_phases} shows the increasing area covered within a given fraction of the maximum intensity measured on the screen, for all the explored painting patterns. For configuration e), a finer grid was used to map the light intensity. %If the larger grid is used, the area covered within a factor of 1/5 increases from 79.8\% to 98.1\% for configuration d) and e), respectively.

\begin{figure}[t!]
\centering
\hspace{-5mm}
\includegraphics[valign=t,scale=0.35]{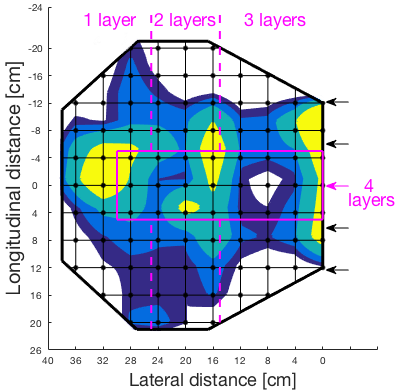}
\includegraphics[valign=t,width=0.3\columnwidth]{legend.pdf}
\caption{Light-intensity contour map of the octagonal screen (color online). The painting pattern is the same as for the configuration in Fig. \ref{3_2_1_ex_layer}, with the addition of the Winston cone on the PMT. Black arrows indicate the light-injection edge. The color code is the same as in Fig.\ref{rectangular_contour_maps}.}
\label{maps_Winston_cone}
\end{figure}

Once satisfying results were achieved in terms of light homogeneity using the brush painting, we decided to mount the Winston cone on the PMT, as in the real camera, to assess its impact on the measurements. Results are shown in Fig.~\ref{maps_Winston_cone}. From a comparison with the map in Fig.~\ref{3_2_1_ex_layer}, the presence of the Winston cone accentuates the impact of the inhomogeneities on the screen. The Winston cone collects light from a defined solid angle preventing the collection of photons coming from further regions of the screen at large angles. The amount of area covered within a factor of 1/5 from the maximum light intensity measured on the screen drops from 86.4\% to 66.4\%.

For configuration a), the simplest one, we decided to test another painting technique in order to see whether a better light homogeneity could be achieved. This technique relies on an air brush filled with pure paint. We immediately realized that the lack of a robotic system controlling the distance and the time spent over each region of the screen makes it difficult to control the amount of applied paint. Moreover, sometimes larger drops of paint were ejected, compromising the final homogeneity. We abandoned this painting method because it did not prove to be a valid alternative to the brush painting.

\begin{figure}[t!]
\centering
\includegraphics[scale=0.2]{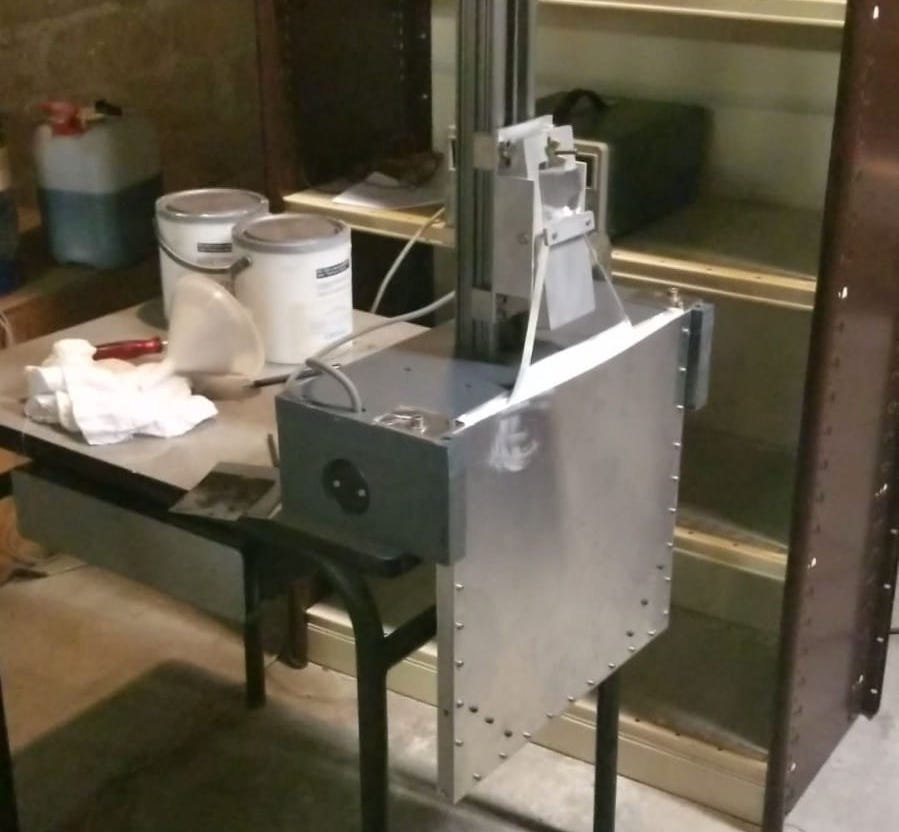}
\caption{Dip-coating painting method. The octagonal screen is immersed in a slim tank 2 $\times$ 44 $\times$ 48 cm$^3$, filled with diluted reflective-paint. To ensure the same deposit of paint over the entire surface, the screen is lifted at a constant velocity of 2 mm/s by a motorized system.}
\label{dipcoating}
\end{figure}

\begin{figure}[t!]
\centering
\hspace{-3mm}
\includegraphics[valign=t,width=0.7\columnwidth]{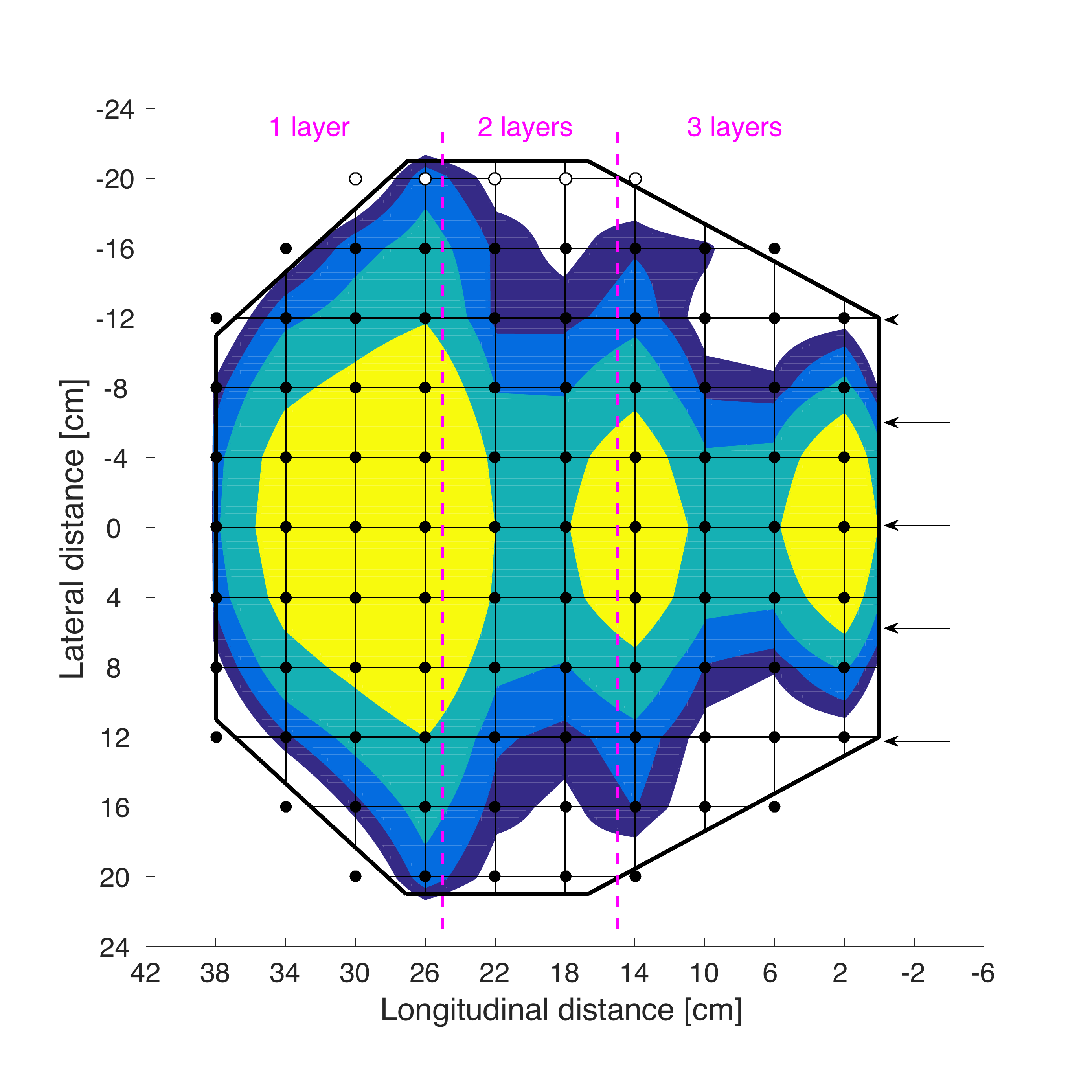}
\includegraphics[valign=t,width=0.3\columnwidth]{legend.pdf}
\caption{Light-intensity contour map (color online). Light is injected into the right edge of the screen (black arrows). Measurements (black dots) were taken on a grid 4 cm $\times$ 4 cm with a Winston cone. The unmarked points outside the border of the screen are set to 0. Because of motor range-limitations, measurements corresponding to the white dots were not taken, and set equal to their symmetric counterparts with respect to the lateral distance $=0$. Magenta dotted lines refer to the painting pattern. The color code is the same as in Fig. \ref{rectangular_contour_maps}.}
\label{oct_ContourMap}
\end{figure}

At this stage, another painting method was tested since it was evident that the brush-painting technique did not permit to deposit the same amount of coating in all regions of the screen. Indeed, a technique that enables the control of the thickness of the paint layers is needed for the pre-production and production phases, in order to ensure a good reproducibility of the performance of each system that will be mounted on several telescopes. For this aim, a dedicated painting method (afterwards referred to as dip-coating) was developed. The screen is immersed in a slim tank containing diluted paint, and then lifted up by a motorized system with a constant speed of 2 mm/s (Fig. \ref{dipcoating}). The first trial, by using pure paint, ended up with a too thick layer of paint and the presence of lumps. Hence, the paint has been progressively diluted with water, until an optimal ratio (80\% paint, 20\% water) was achieved for a homogeneous coverage of the screen. This technique revealed a double advantage: (i) it enables a simplification of the final painting pattern with respect to the brush-painted solution as the extra layer of configuration e) (between 0 and 30 cm along the longitudinal axis, and between -5 and 5 cm along the lateral distance) is no longer necessary; and (ii) it improves the light homogeneity, by extending the amount of area covered within a factor of 1/5 above 80\% of the screen surface. Results are shown in Fig.~\ref{oct_ContourMap}.

\begin{figure}[hb!]
\hspace{-5mm}
\includegraphics[width=1.07\columnwidth]{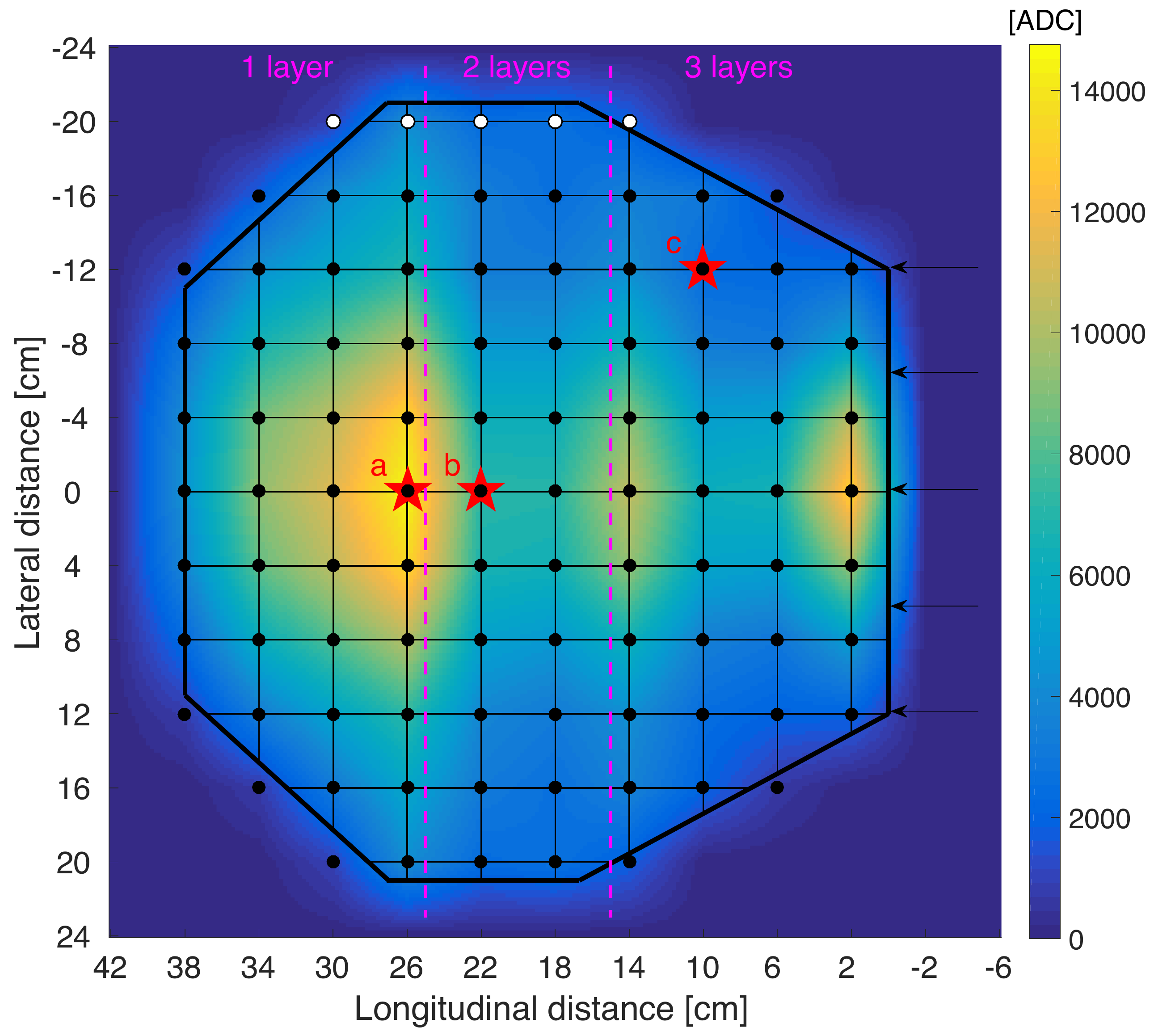}
\caption{Light intensity map (color online). Measurements are the same as in Fig.~\ref{oct_ContourMap}, and were acquired with the Winston cone mounted. Magenta dotted lines refer to the painting pattern. Black arrows indicate the light-injection edge. The three red stars indicate the points where the SPE spectra were acquired (see text and Fig. \ref{spe} for details).}
\label{oct_LightMap}
\end{figure}

%%%%%%%%%%%%%%%%%%%%%%%%%%%%%%%%%%%%%%%%%%%%%%%%%%
\section{Characterization of the final screen}
\label{Optical_prop}

\begin{figure*}[h!]
\centering
\hspace{-5mm}
\subcaptionbox{\label{Arr_time_map}}{\includegraphics[width=1.0\columnwidth]{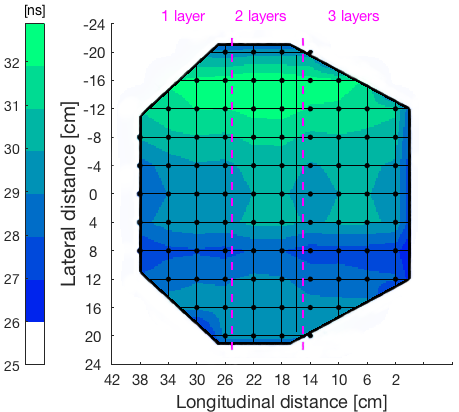}}
\hspace{5mm}
\subcaptionbox{\label{FWHM_map}}{\includegraphics[width=1.0\columnwidth]{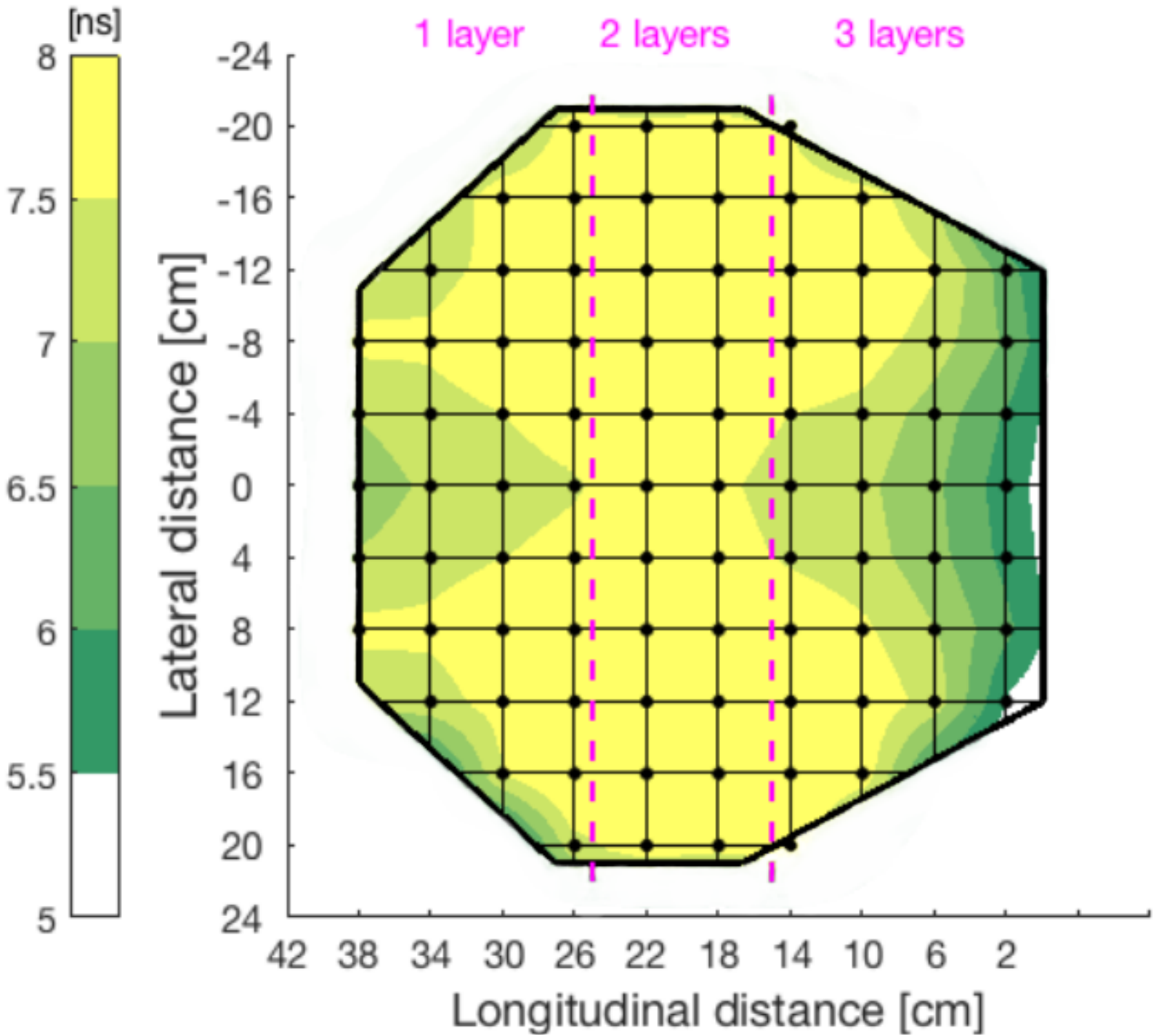}}
\caption{Arrival time map (a), and FWHM map (b) of the signal of the octagonal screen (color online). Light is injected into the right edge of the screen. Measurements (black dots) were taken through on a grid 4 cm $\times$ 4 cm. Magenta dotted lines refer to the painting pattern. The small features close to the borders are due to the interpolation scheme.}
\label{timing}
\end{figure*}

%\begin{figure*}[t!]
%\centering
%\hspace{-5mm}
%\subcaptionbox{\label{oct_LightMap}}{\includegraphics[width=1.07\columnwidth]{Oct_dipcoating_LightMap.pdf}}
%\hspace{5mm}
%\subcaptionbox{\label{oct_ContourMap}}{\includegraphics[width=0.95\columnwidth]{Oct_dipcoating_ContourMap.pdf}}
%\caption{Light intensity map (left) and contours map (right). Light is injected into the right edge of the screen (corresponding to $x=0$). Measurements (black dots) were taken on a grid 4 cm $\times$ 4 cm. The unmarked points outside the border of the screen are set to 0. Because of motor range-limitations, measurements corresponding to the white dots were not taken, and set equal to their symmetric counterparts with respect to the lateral distance $=0$. Magenta dotted lines refer to the painting pattern. In the contour map, the color code is the same as in Fig. \ref{rectangular_contour_maps}. The three red stars indicate the points where the SPE spectra were acquired (see text for details).}
%\label{Optical_properties}
%\end{figure*}

The design process described in Sec. \ref{ReD_activity} led to the choice of an octagonal-shape screen. It is painted with three layers of paint towards the mirrors and 1 layer of paint on the edges, with the exception of the three furthest edges. The development of the dip-coating application method enabled an improvement of the light homogeneity. As a matter of fact, with this method, the painting-pattern towards the focal plane is simplified, receding from configuration e) to configuration d) (see Sec.~\ref{octagonal}), and the additional longitudinal layer is no longer necessary. Even with the Winston cone mounted on the PMT, the light intensity measurements improves significantly both in terms of area covered within a given fraction of the maximum light intensity, and in terms of symmetry with respect to the central longitudinal axis. Figure \ref{oct_LightMap} shows the light intensity map measured over the screen. The percentages of the area covered within a given factor of the maximum light intensity measured on the screen are reported in Table \ref{percent_cov_area}.

\begin{table}[]
\center
\begin{tabular}{|c|c|c|}
\hline
Fraction  & \begin{tabular}[c]{@{}c@{}c@{}}Area {[}\%{]}\end{tabular} & \begin{tabular}[c]{@{}c@{}c@{}}Area {[}cm$^2${]}\end{tabular}\\
\hline
1/2             & 21.9             & 293        \\ \hline
1/3             & 52.3             & 700        \\ \hline
1/4             & 68.8             & 921        \\ \hline
1/5             & 82.2             & 1100      \\ \hline
\end{tabular}
\caption{Percentage of the screen area covered within a given factor of the maximum light intensity.}
\label{percent_cov_area}
\end{table}

A study of the signal timing in terms of arrival time and FWHM, was performed in high-intensity regime on the final configuration (using a PMT equipped with the Winston cone) by acquiring $\sim$30,000 waveforms at each point of a grid 4 cm $\times$ 4 cm over the entire screen. The corresponding maps are shown in Fig. \ref{timing}. The arrival time (defined as the time at the maximum amplitude) of the signal increases with the longitudinal distance, reaching its maximum on the wings of the screen. The maximum delay in the arrival time is $<6$ ns. The FWHM is larger in the wings, most likely in the regions collecting photons reflected from several parts of the screen.
The FWHM of the acquired signal is comparable with the typical duration of a pixel signal arising from a $\gamma$-ray event,\footnote{The time difference between the first and last photo-electrons deposited in a pixel of the camera from a shower at a zenith angle of $20^\circ$ shows a median of ${\sim}2\,$ns for a primary $\gamma$-ray at 100\,GeV and goes up to ${\sim}8\,$ns at 10\,TeV. Shower-to-shower fluctuations induce an rms spread in these estimates on the order of the mean.} ranging from 4.9 ns to 8.3 ns, with a median of 6.6\,ns.

\begin{figure*}[t!]
\centering
\hspace{-5mm}
\subcaptionbox{\label{spe_bright}}{\includegraphics[width=0.7\columnwidth]{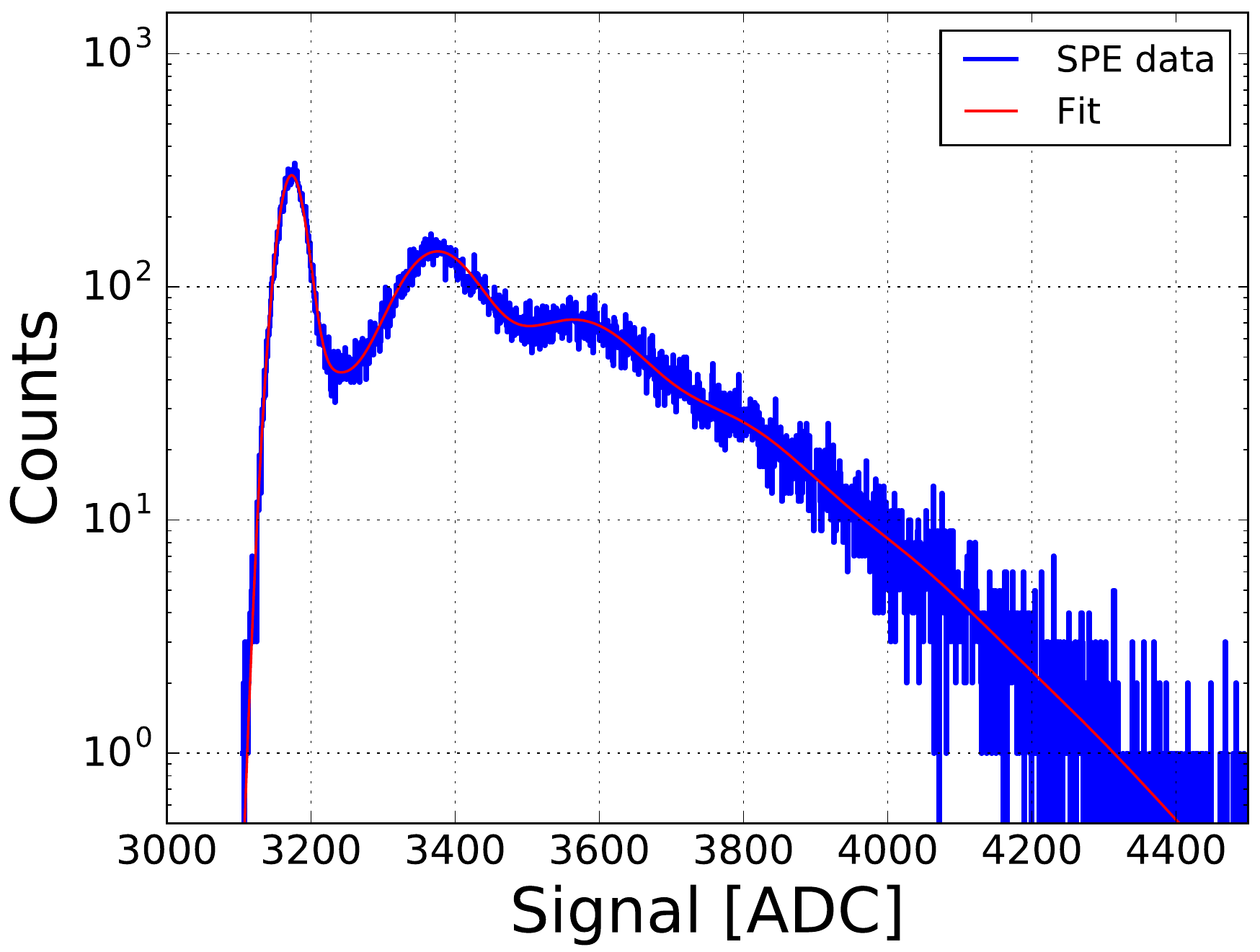}}
%\hspace{2mm}
\subcaptionbox{\label{spe_med}}{\includegraphics[width=0.7\columnwidth]{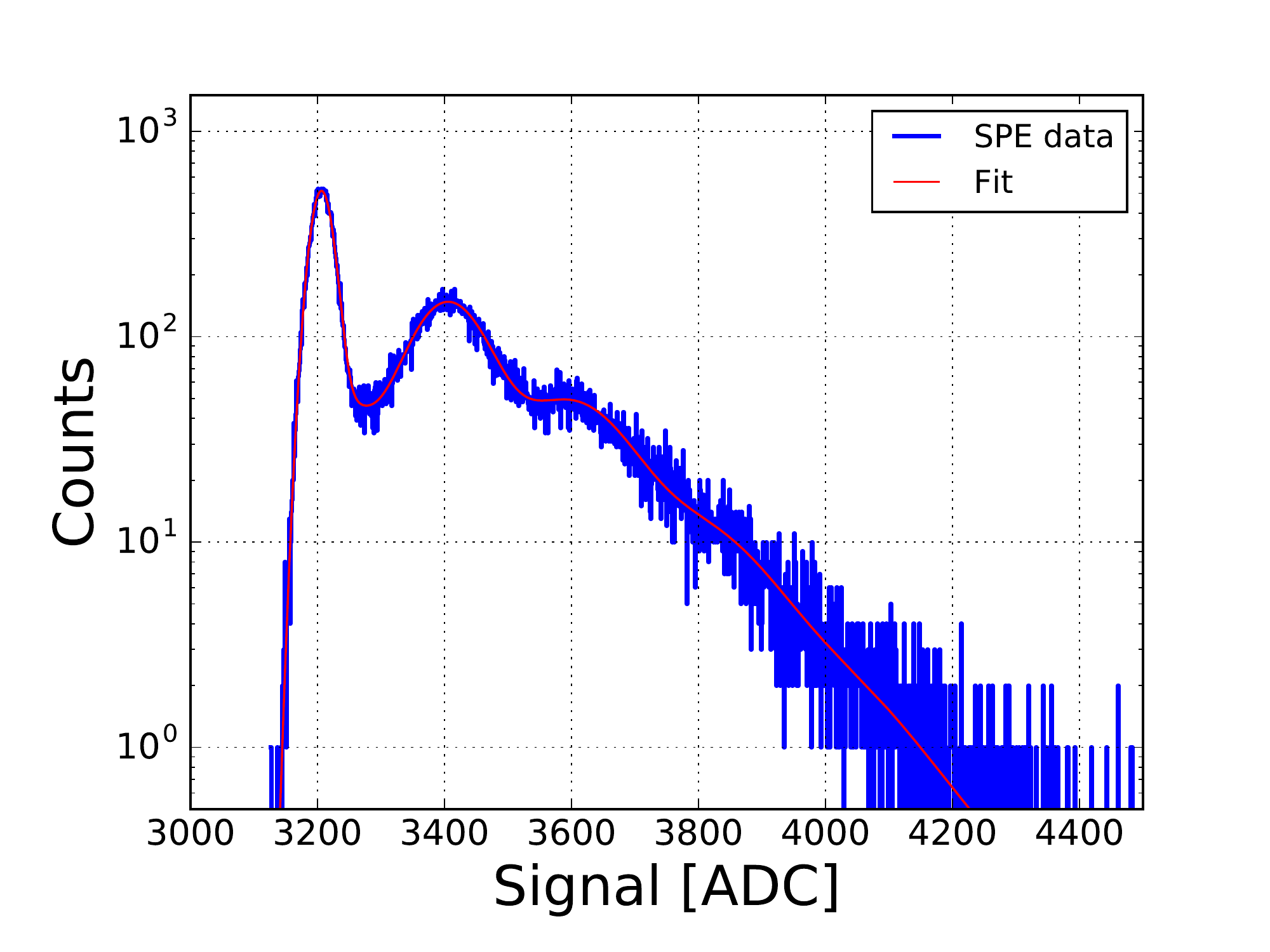}}
%\hspace{2mm}
\subcaptionbox{\label{spe_dark}}{\includegraphics[width=0.7\columnwidth]{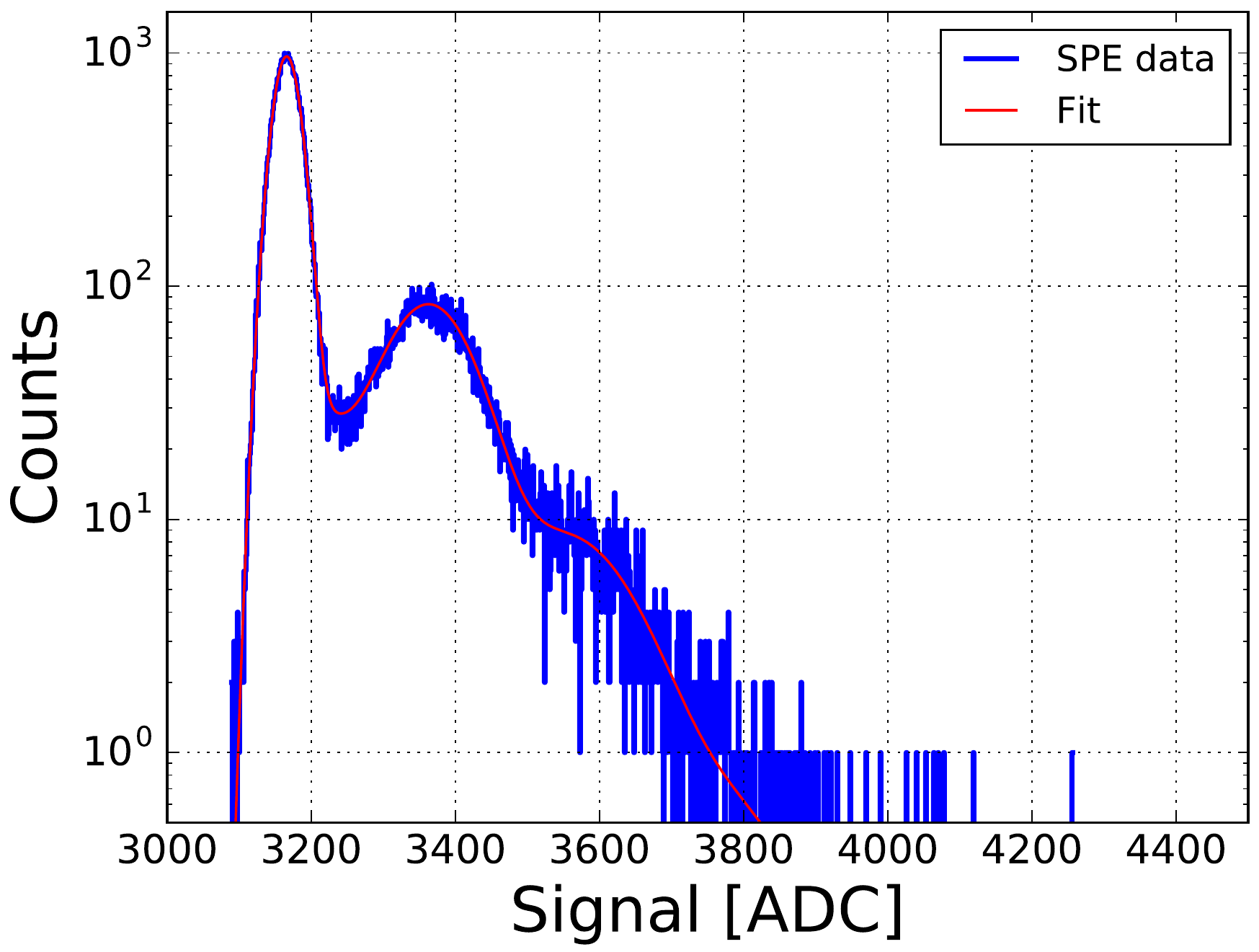}}
\caption{SPE spectra recorded at three different light intensity points of the screen: (a) high, (b) medium, and (c) low. The corresponding positions on the screen are marked in Fig. \ref{oct_ContourMap}. The spectra were acquired by using all 12 LEDs at 12.5 V, with the filter OD=2 mounted in the flasher, and setting the PMT high-voltage at 1200 V.}
\label{spe}
\end{figure*}

The usable area per scan for the SPE acquisition can be evaluated according to the quality of the spectrum and its statistical uncertainties.
Three SPE spectra, containing 60,000 events, whose charge was integrated within a 20 ns window, were taken at different points of the screen: at a high-, at a medium-, and at a low-intensity point, respectively, as shown in Fig.~\ref{oct_LightMap}.
% bright (14753 ADC), in a medium (7117 ADC), and in a low (2421 ADC) intensity point, respectively.
A probability distribution function (PDF) resulting from the convolution of the PDF of the pedestal and $n$ single photoelectron PDF, was then fit to each spectrum.\footnote{The fit was performed with the open-source software Calin, available at \url{https://github.com/llr-cta/calin}} The pedestal PDF is modeled by a Gaussian function, while the SPE PDF is modeled by two Gaussians to take into account the presence of a low-charge event population \cite{Caroff2019}. The latter could be due to electrons that, migrating from the first to the second dynode of the PMT, do not follow an ideal trajectory producing a loss of electrons at the first amplification step, or to photons that produce an electron at the first dynode instead of at the photocatode. This kind of events lead to the formation of a low charge component in the SPE PDF. The SPE spectra are shown in Fig. \ref{spe}. The gain derived at each position is $166.0 \pm 0.6_{\rm stat}$, $165.9 \pm 0.5_{\rm stat}$, and $166.5 \pm 0.7_{\rm stat}$ ADC/photoelectron, for corresponding $I/I_{max}$ light intensity of 1.0, 0.6, and 0.2, respectively (that corresponds to 1.57 $\pm$ 0.09$_{\rm stat}$, 1.04 $\pm$ 0.07$_{\rm stat}$, and 0.33 $\pm$ 0.05$_{\rm stat}$ photoelectrons, respectively).\footnote{The light intensity corresponds to the mean number of photoelectrons of the distribution.} The estimation of the gain does not seem to be affected by the differences in emission over the screen. This means that the entire screen surface could be used for the PMT gain estimation. In-situ measurements in the camera will enable an assessment of the full capability of the SPE system.

%%%%%%%%%%%%%%%%%%%%%%%%%%%%%%%%%%%%%%%%%%%%%%%%%%
\section{Summary}
\label{conclusions}

The SPE system that we developed has a dual purpose: the study of the optical PSF, both on- and off-axis, and the estimation of the gain of the whole photo-detection chain. A movable target is needed for PSF studies. The addition of a light source enables a robust estimation of the gain, which is obtained from the SPE spectrum acquired with each PMT.
The SPE calibration system exploits light pulses injected into a PMMA screen covered by a special pattern of reflective paint.

In order to scan the entire camera, 80 repositionings of the screen are needed. Operating the flasher at a frequency of $\mathcal{O}(100)$ Hz, and acquiring $\mathcal{O}(10^4)$ events for each SPE, the estimated amount of time for a full camera scan is less than three hours. The scan can be performed during daytime provided a lightproof camera enclosure.

In this paper, we reviewed the design process of the screen and characterized its optical performance. In the design process, the following elements have been investigated: (i) the geometry of the screen; (ii) the coating type; (iii) the coating application process, and (iv) coating patterns on the screen.

The final design satisfies the requirements specified for the NectarCAM project, and consists of an octagonal screen, painted with a Bicron reflective paint (Saint-Gobain BC-620) using the dip-coating method. The painting pattern that optimizes the light homogeneity consists of 3 layers in the first 15 cm from the injection edge, 2 layers between 15 and 25 cm, and 1 layer over the rest of the screen. The percentage of area covered within a factor of $1/5$ from the maximum light-intensity measured over the screen is larger than 80\%. The measurements of the arrival time and of the FWHM of the signal, performed with a first version of the prototype, are fully satisfactory, showing a maximum delay of $6$ ns in the arrival time, and a median FWHM of 6.6 ns. Moreover, preliminary SPE spectra, acquired at different brightness points, show that the entire screen surface can be used to determine the gain of the NectarCAM photodetection chain.

The SPE system was integrated and validated with dedicated tests inside the NectarCAM camera prototype mounted on the MST prototype structure in Berlin-Adlershof.
Further studies will be dedicated to the study of the gain of the photo-detection chain as a function of the night-sky-background level and to the feasibility of integrating the SPE system in other CTA cameras. 

\vspace{5mm}
\section*{Acknowledgements}
\noindent This work was conducted in the context of the NectarCAM Project of CTA. The authors are grateful for fruitful collaborations with colleagues from the CTA Consortium, in particular from the MST Structure and NectarCAM Projects.
The authors acknowledge support from P2IO LabEx (ANR-10-LABX-0038) in the framework ``Investissements d'Avenir'' (ANR-11-IDEX-0003-01) managed by the Agence Nationale de la Recherche (ANR, France). Furthermore, we thank F. Cassol for feedback on the expected duration of $\gamma$-ray showers in the camera.
%B.B.~acknowledges support from P2IO LabEx (ANR-10-LABX-0038) in the framework ``Investissements d'Avenir'' (ANR-11-IDEX-0003-01) managed by the Agence Nationale de la Recherche (ANR, France).

%%%%%%%%%%%%%%%%%%%%%%%%
%   BIBLIOGRAPHY

%% `Elsevier LaTeX' style
\bibliographystyle{elsarticle-num}
%\bibliographystyle{elsarticle-harv}
%\bibliographystyle{elsarticle-num-names}
%\bibliographystyle{model1-num-names}
%bibliographystyle{plain}

%\bibliography{bibliography_full}
\bibliography{bibliography}

%\end{linenumbers}
\end{document}